\documentclass{elsart}
\usepackage[english]{babel}
\usepackage{latexsym}
\usepackage{fancyhdr}
\usepackage{stmaryrd}
\usepackage{amsmath,amssymb,graphicx}
\usepackage[numbers]{natbib}
\journal{Energy Economics.}
\def\cite{\citet}

\newcommand{\Eb}{\mathbb E}

\newcommand{\Qb}{\mathbb Q}
\newcommand{\Pb}{\mathbb P}
\newcommand{\dd}{{\rm d}}
\begin{document}
\begin{frontmatter}


\title{
Arbitrage free cointegrated models in gas and oil future markets}
\author[GDF]{Gr\'egory Benmenzer},
  \ead{gregory.benmenzer@free.fr}
\author[Grenoble]{Emmanuel Gobet},
  \ead{emmanuel.gobet@imag.fr}
\author[GDF]{C\'eline J\'erusalem}
  \ead{celine.jerusalem@gazdefrance.com}
\address[GDF]{Gaz de France, Research and Development Division,
361 Avenue du Pr\'esident Wilson - B.P. 33, 93211 Saint-Denis La
Plaine cedex, FRANCE}
\address[Grenoble]{Laboratoire Jean Kuntzmann, Universit\'e de Grenoble and CNRS, BP 53, 38041 Grenoble cedex 9, FRANCE}
\begin{abstract}
In this article we present a continuous time model for natural gas and crude oil future prices. Its main feature is the possibility to link both energies in the long term and in the short term. For each energy, the future returns are represented as the sum of volatility functions driven by motions. Under the risk neutral probability, the motions of both energies are correlated Brownian motions while under the historical probability, they are cointegrated by a Vectorial Error Correction Model. Our approach is equivalent to defining the market price of risk. This model is free of arbitrage: thus, it can be used for risk management as well for option pricing issues. 
Calibration on European market data and numerical simulations illustrate well its behavior.
\end{abstract}

\begin{keyword}
future prices, natural gas, crude oil, cointegration, Vectorial Error Correction Model, arbitrage free modelling.
\end{keyword}

\end{frontmatter}

\ack{The second author is grateful to Gaz de France for its support. Furthermore, we thank Antoine J\'erusalem for his advice.}

\section*{Introduction}


The need to model future prices of gas and oil simultaneously to optimize energy portfolios is now very present. Indeed an energy portfolio could be invested into several energy markets which could interfere with each other. For example, a firm which detains a supply contract will need to model gas and oil evolution for an optimal risk management.

With the energy market liberalisation over the last decade, one may think that gas and oil prices are decoupled, but actually several statistical studies tend to prove that prices are cointegrated. For statistical evidences of cointegration and economic explanations, we refer for instance to the articles by \cite{pana:rutl:07} and \cite{asch:osmu:sand:06} for the UK and European markets, and to the article by \cite{bach:grif:06} for the US market. The dependence between gas and oil prices could be economically explained with gas long term contracts, which still represent the majority of supply in European gas and whose prices are indexed on oil and oil products prices. This indexation creates a structural link between prices of both energies.


In the previously cited references, one proposes econometric models (in discrete time) for gas/oil prices, which are coherent with market data and which accounts well for the interdependence of prices. They are useful for some risk management purposes, such as Value at Risk measurements. However, as soon as we have to consider energy contracts and related pricing/hedging issues, different models emerge in order to be consistent with the arbitrage free theory: they are such that forward contracts are martingales under risk neutral probabilities (see \cite{musi:rutk:98}). In the following and to simplify our presentation, we identify future prices (given by quotations data) and forward prices (given by models), which is correct if interest rates are deterministic for instance. Usual factor models for the spot and forward contracts on a given energy are written as
\begin{equation}
  \label{eq:forward}
  \frac{\dd F(t,T)}{F(t,T)} = \sigma(t,T) \Sigma  \dd B_t
\end{equation}
where
\begin{itemize}
\item $F(t,T)$ is the forward contract quoted in $t$ and delivered in $T$,
\item $\sigma(t,T) = \left( \sigma_1(t,T), \sigma_2(t,T), \ldots, \sigma_n(t,T) \right)$ is a row vector of normalized volatility functions of forward returns,
\item $n$ is the number of risk factors identified through a PCA (Principal Component Analysis) of forward returns,
\item $B_t = \left( B_t^1, B_t^2, \ldots, B_t^n \right)^*$ are independant Brownians motions under risk neutral probabilities (here ${}^*$ stands for the transposition),
\item $\Sigma$ is a $n\times n$ matrix, equal to the square root of a variance-covariance matrix.
\end{itemize}
See the works by \cite{Geman}, \cite{ClewlowStrickland} among others.

The volatility functions describe the {\em shifting}, the {\em twisting} and the {\em bending} (for \cite{ClewlowStrickland}). In \cite{Brooks}, they describe the {\em level}, the {\em slope}, the {\em curvature} (LSC model). R.~Brooks uses this model for natural gas contracts and gives an explicit form for these functions :
$$\sigma_1^e (T-t)=1,$$
$$\sigma_2^e (T-t)=e^{-\frac{T-t}{\tau_1^e}},$$
$$\sigma_i^e (T-t)=\frac{T-t}{\tau_{i-1}^e} e^{-\frac{T-t}{\tau_{i-1}^e}} \mbox{ for } i \geq 3.$$
The model is linear in the parameters ($\Sigma$ and $(\tau_i^e)_i$), thus ordinary least squares regression is applied to estimate them.

For crude oil contracts, analogous models could be set up, with different Brownian motions. These ones can be correlated to those of gas models. Within this approach, it appears that volatility functions adjust well for each energy. Nevertheless, long term dependences are poorly modeled. The variance-covariance matrix (related to $\Sigma$) induces relevant marginal distributions of the returns but unfortunately unrealistic joint price distribution. 
For example, when simulating the model one often obtains a growing prices' scenario for gas and a decreasing one for crude oil (see Figure \ref{BrooksSimu} in Section \ref{section:test}). It is fundamental to note that the previously decribed models \eqref{eq:forward} are written under the risk-neutral probability (denoted by $\Qb$ in the sequel), which is suitable for pricing/hedging issues, while the long-term dependence (given by an econometric cointegration analysis) holds under the historical (or physical) probability (denoted by $\Pb$). To accommodate both features (long term dependences on the one hand; stochastic returns described by volatility functions on the other hand), a natural idea consists in suitably modeling the market price of risk $(\lambda_t)_t$, making the connection between historical and risk-neutral worlds. This is the main contribution of our work. Details are given in Section \ref{section:model}. Thus, our model cointegrates gas and oil prices while being coherent with the pricing by arbitrage. In the following, we mainly focus on natural gas and crude oil.
 
We now mention a similar approach to ours. In his PhD thesis, Steve Ohana aims at modeling spot and forward contracts for two cointegrated energies, but in discrete time (see \cite{Ohana}). He handles the case of US natural gas and crude oil market.
For each energy $e$ ($e=g$ for gas, $e=c$ for crude), the model is given by (under $\Pb$)
$$ \frac{\Delta F^e(t,T)}{F^e(t,T)} = e^{-k_e (T-t)} \Delta X_t^e + \Delta Y_t^e$$
where 
$\frac{1}{k_e}$ is the characteristic time of the short term shock. 
The vector $(\Delta X_t^g, \Delta X_t^c,\break \Delta Y_t^g, \Delta Y_t^c )$ has a drift component, equal to a constant plus two terms related to the past and present of the process, and a noise component expressed as independent GARCH processes. The dependence on present is defined by the non-linear link between $(\Delta X_t^g, \Delta X_t^c)$ and the processes $(X_t^g)$ and $(X_t^c)$ and between $(\Delta Y_t^g,\Delta Y_t^c)$ and the processes $(Y_t^g)$ and $(Y_t^c)$.

Finally we mention very recent works, whose main motivations are to analyse the market risk premium $\pi^e(t,T)$ through the specification of the market price of risk $(\lambda_t)_t$. We recall that the market risk premium of the energy $e$ is defined by the difference between forward prices and expected future spot prices (under the physical probability $\Pb$)\footnote{note that the sign of $\pi^e$ is not related to backwardation and contango situation.}:
$$\pi^e(t,T)=F^e(t,T)-\Eb^\Pb(S_T|{\mathcal F}_t),$$
where ${\mathcal F}_t$ stands for the information up to time $t$. \cite{cart:will:07} are dealing with the two factors long term/short term model in the US gas market. \cite{kolo:ronn:07} provide a statistical methodology to estimate constant commodity market prices of risk. \cite{bent:cart:kies:06} make explicit the connection between $\lambda$ and $\pi^e$ through the market players' risk preferences. All these works focus on an energy at a time. Our approach is rather different because we specify the risk market price in order to design long term dependences between gas and oil. Regarding the market risk premium, with a suitable choice of $(\lambda_t)_t$ we propose to adjust it to 0 at time 0 because it is often a requirement of practitioners. However, it could be adjusted to any value, to generate positive or negative market risk premia.

The model presented in the following (Section \ref{section:model}) is mainly inspired by the works of \cite{Brooks} and \cite{Ohana}. We use a cointegration approach and we design a Vectorial Error Correction Model (VECM). This type of model is shortly discussed in Section \ref{section:vecm}. Compared to Ohana's model, a continuous time setting is used, which is compatible with the arbitrage free theory. Note that the cointegration and VECM are now standard tools in energy markets: see for instance \cite{deva:wall:99} for spot electricity prices, \cite{serl:herb:99} for US gas/oil prices. 
Finally, in a third part, we make numerical simulations to test the model.


\section{Cointegration and Vectorial Error Correction Model (VECM)}
\label{section:vecm}

Expressed in 1987 by \cite{EngleGranger}, the cointegration concept characterizes the fact that in long term, a specific combination of non-stationary processes could be stationary. A time series vector $\{y_t=(y^1_t, \ldots, y^n_t):t\in\mathbb{N}\}$ is said cointregrated if each series $(y^i_t)$ is integrated with an order of integration equal to 1, and if some linear combination of the series $\alpha\cdot y_t$ is stationary. This linear combination is the long term equilibrium between each component of the vector $y_t$.

\begin{figure}[htbp]
\begin{center}
\includegraphics[height=6cm,width=6cm]{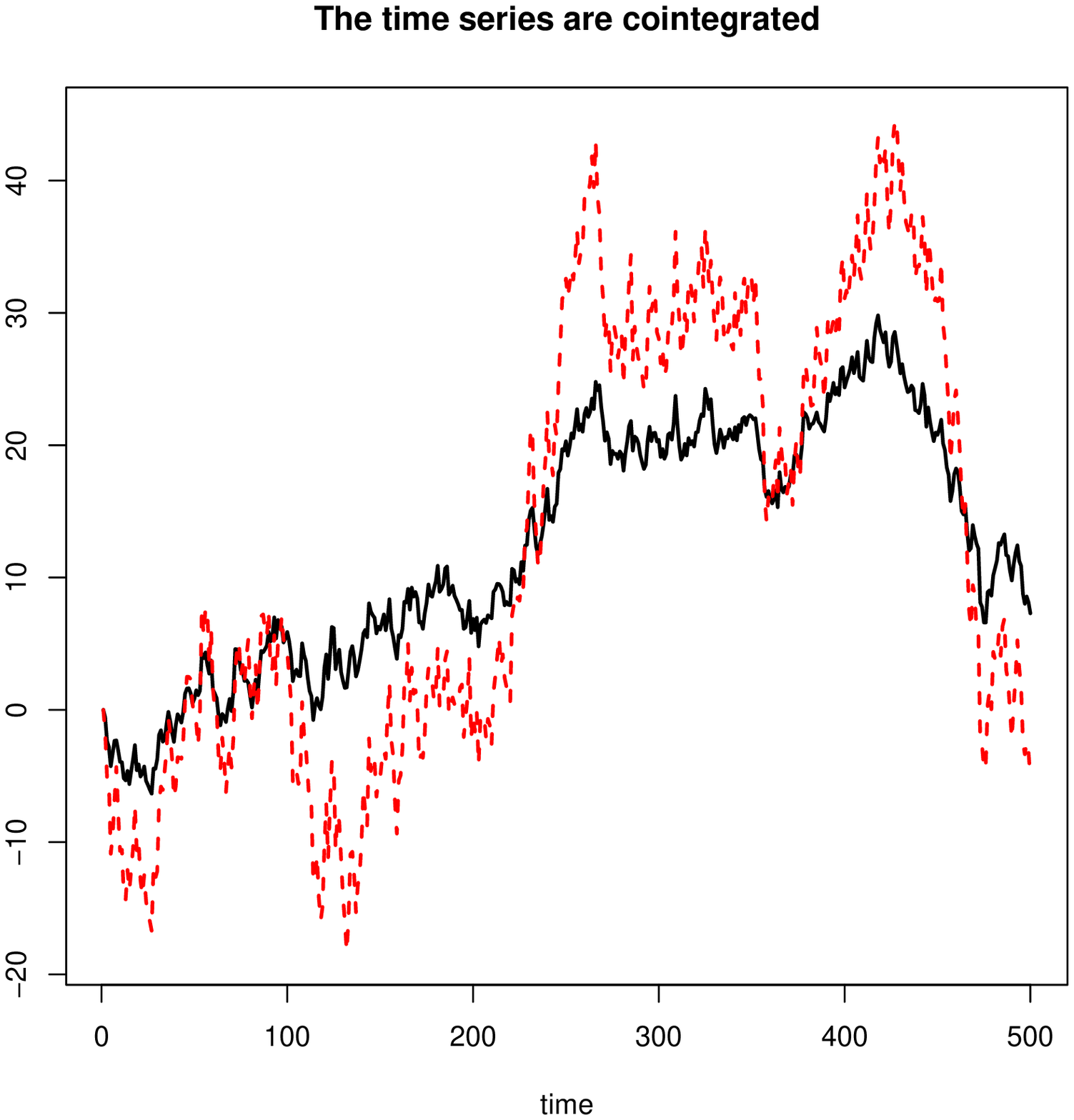}
\includegraphics[height=6cm,width=6cm]{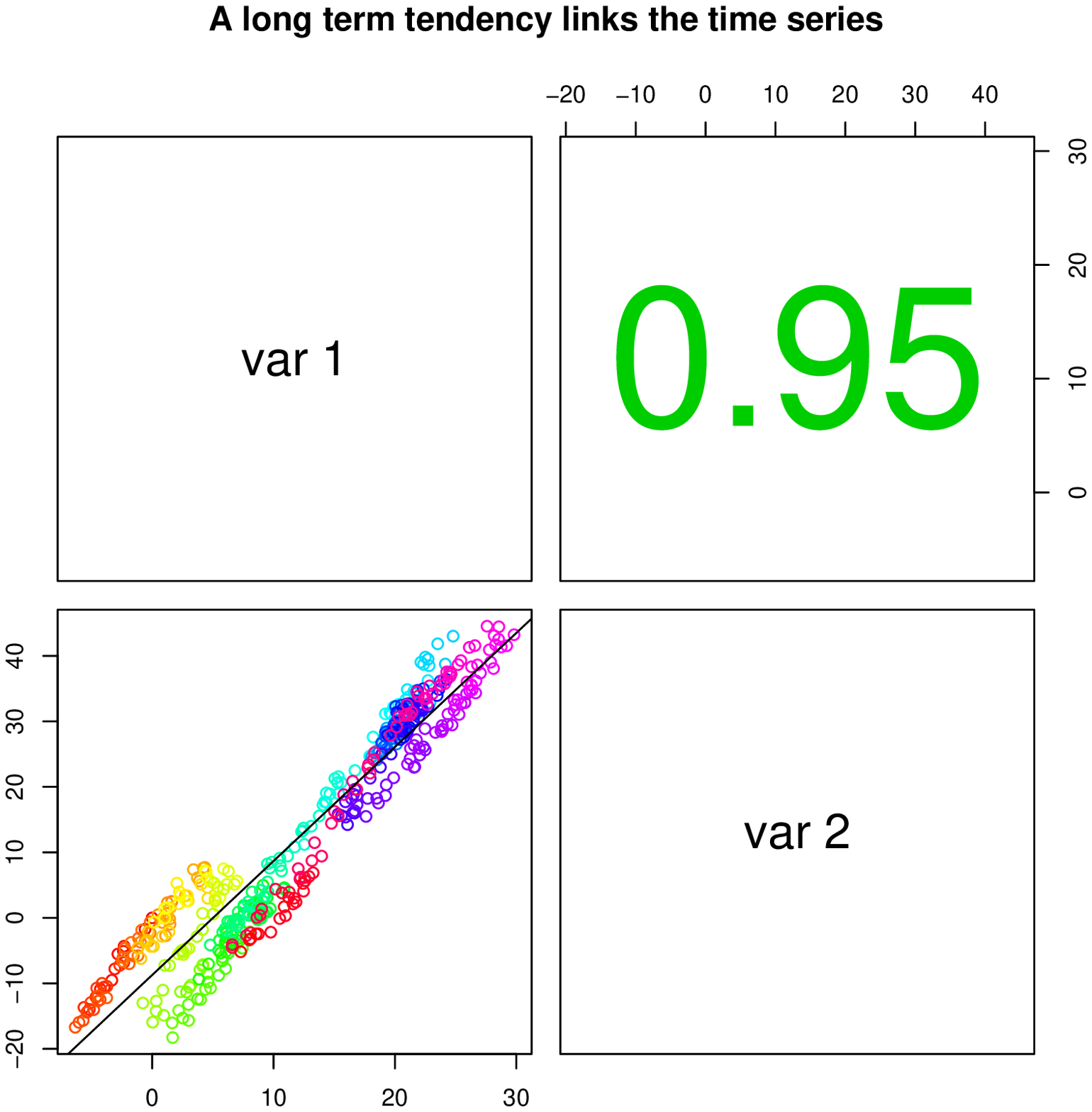}
\end{center}
\caption{Example of cointegrated time series.\newline
\small{ \textit{The graph on the right side allows us to see the scatter plot of both processes and the correlation between their trajectory. The size and the color of the font change according to the value of the correlation: the higher the correlation is, the larger the font is.}}	
\label{ExCoint}}
\end{figure}

We can give a simple example based on \cite{Hamilton}:
\begin{equation} \label{EquCoint}
\left\{
\begin{array}{l}
\Delta y^1_t= \Delta W_t^1,\\
y_t^2 = 2 y_t^1 + \Delta W_{t-1}^2,\\
y_0^1=y_0^2=0.
\end{array}
\right.
\end{equation}
Possible simulation of this model is given on Figure \ref{ExCoint}. If $(\Delta W_t^1,\Delta W_t^2)$ are uncorrelated white noise process, $(y^1_t,y^2_t)$ are cointegrated.

The Equation (\ref{EquCoint}) can be rewritten :
\begin{equation} 
\left\{
\begin{array}{l}
\Delta y^1_t= \Delta W_t^1,\\
\Delta y_t^2 = 2 y_t^1 - y_t^2 + 2 \Delta W_t^1 + \Delta W_{t}^2.
\end{array}
\right.
\nonumber
\end{equation}
This is the form of a simple VECM (Vectorial Error Correction Model). More generally, if a time series vector $\{y_t=(y^1_t, \ldots, y^n_t):t\in\mathbb{N}\}$ is cointegrated, the following relation could model the cointegration relation between the components of $y_t$:
$$\Delta y_t= \Pi y_t + \Sigma  \Delta W_t. $$
The matrix $\Pi$ represents the long term relation while the matrix $\Sigma$ stands for the short term behavior.

We extend this notion to a continuous time framework, by writing
\begin{equation}
  \label{eq:vecm:continu}
  \dd y_t= \Pi y_t\dd t + \Sigma \dd W_t
\end{equation}
where $(W_t)_t$ is a vector of independent Brownian motions.

We point out that $(y_t)_t$'s dynamics identifies with that of a multidimensional Ornstein-Uhlenbeck process, with possibly a degenerate matrix $\Pi$. In fact, several cases are possible:
\begin{itemize}
\item If the rank of $\Pi$ is full ($rk(\Pi)=n$), motions $y_t$ may be stationary processes and $n$ cointegrating relations exist. 
\item If the rank of $\Pi$ is zero ($rk(\Pi)=0$), motions $y_t$ are not cointegrated and not stationnary.
\item If the rank of $\Pi$ is $r$ with $0<r<n$, two $(n \times r)$ matrices $\alpha$ and $\beta$ exist such as $\Pi=\alpha \beta^*$. The $r$ linear independent columns of $\beta$ are cointegrating vectors and the elements of $\alpha$ determine the speed of the long term return. Motions $y_t$ are cointegrated and non-stationary. 
\end{itemize}
For more details, see \cite{Pfaff}.

\section{The model}
\label{section:model}
To simplify our presentation, we assume that the prices of energy contracts are not correlated with the currencies\footnote{To avoid extra contributions related to currency volatilities in the model.} and that interest rates are deterministic\footnote{This makes future and forward prices equal.}. In the following, we choose to model gas and oil forward prices as the sum of volatility functions, which are driven by cointegrated motions. These motions depend on a deterministic function which can be adjusted to centre prices on the initial forward curve. The cointegrated term is related to the market price of risk.
\subsection{Derivation of the model}
We first write the dynamics of the forward prices' return for both energies under the risk neutral probability\footnote{Under our assumptions on interest rates, all the forward risk-neutral probabilities coincide (and are equal to $\Qb$). Thus for any $T$, $(F^e(t,T))_t$ is a martingale under $\Qb$.} $\Qb$ (this is the same dynamics as Brooks' model):
\begin{equation}
  \label{eq:forward:2}
\left\{
\begin{array}{l}
\frac{\dd F^g(t,T)}{F^g(t,T)} = \sigma^g (T-t) \dd X_t,\\
\frac{\dd F^c(t,T)}{F^c(t,T)} = \sigma^c (T-t) \dd X_t,\\
\dd X_t=\Sigma \dd B_t,\\
\end{array}
\right.
\end{equation}
where we set
\begin{itemize}
\item $F^e(t,T)$ for the forward price quoted in $t$, delivering at the date $T$ one unit of the energy $e$ ($g$ for natural gas or $c$ for crude oil). 
\item $N^e$ for the factors' number for the energy $e$ (in practice $N^e=3$). This is the number of motions kept in the PCA during the calibration.
\item $B_t=(B_t^1,\ldots,B_t^{N^g+N^c})^*$ for independant $\Qb$-Brownian motions.
\item $\Sigma$ for a non-degenerate $(N^g+N^c)\times (N^g+N^c)$ matrix, used to correlate the returns.
\item $(\sigma_i^e (T-t))_i$ for the set of normalized volatility functions for the energy $e$:
\begin{itemize}
\item $\sigma_1^e (T-t)=1,$
\item $\sigma_2^e (T-t)=e^{-\frac{T-t}{\tau_1^e}},$
\item $\sigma_3^e (T-t)=\frac{T-t}{\tau_{2}^e} e^{-\frac{T-t}{\tau_{2}^e}}.$
\end{itemize}
\item $\sigma^g (T-t)$ and $\sigma^c (T-t)$ for the volatilities of each energy as $(N^g+N^c)$-dimensional row vectors defined by
\begin{align*}
  \sigma^g (T-t)& =  (\sigma_1^g,\ldots,\sigma_{N^g}^g,0,\ldots,0) (T-t),\\ 
\sigma^c (T-t) &=  (0,\ldots,0,\sigma_1^c,\ldots,\sigma_{N^c}^c) (T-t).
\end{align*}
\end{itemize}
$(X_t)_t$ are the so-called $N^g+N^c$ motions of gas and oil: under $\Qb$, we have defined them as non-normalized Brownian motions. In the above definition, the initial value of $X$ is arbitrary because only increments of $X$ have an impact on the dynamics of forward prices (under $\Qb$).

In the working paper of \cite{Brooks}, we can find a complete explanation of the different factors. $\sigma_1$ is the volatility coefficient for the {\em level} of change in the forward curve. Since $\dd F(t,T)$ behaves as $F(t,T) \sigma_1 \dd X_t^1$ when $T-t$ tends to infinity, the first risk function drives the long term volatility. $\sigma_2$ is the volatility coefficient for the {\em slope} of changes in the forward curve, and the other term $\sigma_3$ is the volatility coefficient associated to the {\em curvature}.

Let us go back to a model under the historical probability $\Pb$. For this, we take advantage of the fact that $(B_t)_t$ corrected by the market price of risk $(\lambda_t)_t$ becomes a $\Pb$-Brownian motion, which we denote by $(W_t)_t$ (see \cite{musi:rutk:98}). We have 
$$B_t=W_t+\int_0^t \lambda_s \dd s.$$
$(\lambda_t)_t$ is a $(N^g+N^c)$ dimensional process, which we define as
\begin{equation}
  \label{eq:prime}
  \lambda_t=\Sigma^{-1}[\Pi X_t +\eta_t],
\end{equation}
where
\begin{itemize}
\item $\Pi$ is a $(N^g+N^c)\times (N^g+N^c)$ matrix representing the long term link between the motions $(X_t)_t$,
\item $(\eta_t)_t$ is a $(N^g+N^c)$-dimensional process taken as deterministic in the following. This will play the role of a centring factor.
\end{itemize}
This definition of $(\lambda_t)_t$ is crucial in our model. In this way, $\lambda$ represents the cointegration effect between energies, by putting a long term relation between the motions $(X_t)_t$:
$$\dd X_t=\Pi X_t \dd t +\Sigma \dd W_t+\eta_t \dd t.$$
The term $\Pi X_t$ cointegrates motions of gas and oil. The above equation is a continuous time VECM as the ones described by \eqref{eq:vecm:continu}.

\subsection{Final form of the model}
To sum up, under the historical probability $\Pb$, the dynamics of the forward prices return is:  
\begin{equation}
\label{eq:model}
\left\{
\begin{array}{l}
\frac{\dd F^g(t,T)}{F^g(t,T)} = \sigma^g (T-t) \dd X_t,\\
\frac{\dd F^c(t,T)}{F^c(t,T)} = \sigma^c (T-t) \dd X_t,\\
\dd X_t = \Pi X_t \dd t + \Sigma \dd W_t + \eta_t \dd t,
\end{array}
\right.
\end{equation}
where $W$ is a standard $\Pb$-Brownian motion. Under (spot and forward) risk neutral probabilities, one has $\dd X_t = \Sigma \dd B_t$ with the standard $\Qb$-Brownian motion $B$.

\subsection{Calibration of the model} 
In practice, the available data are related to gas and oil future contracts, and not to derivatives on gas/oil (which are OTC contracts)\footnote{Market data related to energy derivatives would be useful to better calibrate the volatility functions.}. Hence, we perform a statistical calibration. 

The determination of the parameters is done in five steps:
\begin{enumerate}
\item Principal Components analysis (PCA) on the returns (they are computed by $\frac{F^e(t+1\text{ day},T)-F^e(t,T)}{F^e(t,T)}$).
\item Estimation of the parameters $\tau_i^e$ ($i=1,2$) with a nonlinear regression between $1, e^{-\frac{T-t}{\tau_1^e}},\frac{T-t}{\tau_2^e} e^{-\frac{T-t}{\tau_2^e}}$ and the motions $X$ deduced from the PCA.	
\item Reconstruction of the differences $\dd X_t^e$ at each time step with linear regression between the volatility functions vector and the returns vector.
\item Linear regressions between $(\Delta X_t)$ and $(X_t)$ to determine the matrices $\Pi$ and $\Sigma \Sigma^*$. In the matrix $\Pi$, we keep significant elements with a regression subset selection according to the Bayesian Information Criterium (BIC), see \cite{Miller}. 
\item Computation of the function $(\eta_t)_t$.
\end{enumerate}
Only the last step needs to be detailed. We mention first that we have set $X_0=0$ in the previous steps, which is equivalent to shift the (unknown) value of $(\eta_t)_t$.\\
$(\eta_t)_t$ is an additional drift parameter. Usually, it can be identified only over a very long time data set, which is impossible in practice. 

To overcome this undertermination problem, we propose to adjust it so that expected forward prices will fit the initial curve (for each energy):
 $$\Eb^{\Pb} \left(F^e(t,T) \right) = F^e(0,T),\qquad \forall (t,T).$$
This is often expected by practitioners. Note that this adjustment implies a vanishing market risk premium $\pi^e(0,T)$ but any alternative choice would be possible. The model is now rewritten in a more simple manner. We set
$X_t=\tilde X_t+\theta_t$ where $(\theta_t)_t$ solves $\theta_t=\int_0^t (\Pi \theta_s + \eta_s )\dd s$, which is given by $\theta_t=e^{t\Pi} \int_0^te^{-s\Pi}\eta_s \dd s$. Thus, it is equivalent to determine $\eta$ or $\theta$. Equations \eqref{eq:model} consequently become
\begin{equation}
\label{eq:model:bis}
\left\{
\begin{array}{l}
\frac{\dd F^g(t,T)}{F^g(t,T)} = \sigma^g (T-t) (\dd \tilde X_t + \theta'_t \dd t),\\
\frac{\dd F^c(t,T)}{F^c(t,T)} = \sigma^c (T-t) (\dd \tilde X_t + \theta'_t \dd t),\\
\dd \tilde X_t = \Pi \tilde X_t \dd t + \Sigma \dd W_t, \qquad \tilde X_0=0.
\end{array}
\right.
\end{equation}
In the Appendix, we prove that 
\begin{equation}
  \label{eq:factorisation}
  \Eb^{\Pb} \left( \frac{F^e(t,T)}{F^e(0,T)} \right)=
\exp\left(\int_0^t \sigma^e (T-s) \theta'_s \dd s\right)
\Eb_{\theta=0}^{\Pb} \left( \frac{F^e(t,T)}{F^e(0,T)} \right).
\end{equation}
In addition, a closed formula for $\Eb_{\theta=0}^{\Pb} \left(\frac{F^e(t,T)}{F^e(0,T)}\right)$ is given (see \eqref{eq:closed:formula}) which only depends on the already estimated parameters. 
In order to have the left hand side equal to 1, one should have 
\begin{equation}
  \label{eq:choix:theta}
  \sigma^e (T-t)\theta'_t = -\frac{\partial\left[\ln\left(\Eb_{\theta=0}^{\Pb} \left( \frac{F^e(t,T)}{F^e(0,T)} \right)\right)  \right]}{\partial t}
\end{equation}
for any $(t,T)$ and for each energy, with a suitable choice of $\theta$. In our tests on market data, the perfect fit is achieved with less that $0.1\%$ error (see Figures \ref{Theta}).
In that case, $\theta$ is obtained by a least squares optimisation routine applied to the difference of the two sides of \eqref{eq:choix:theta}.

Note that when $\Pi$ is zero, the forward prices are already centred on the initial forward curve if $\theta\equiv 0$: no adjustment of $\theta$ is needed. $X$ are just non standard Brownian motions and the model comes down to the one given by Brooks.

\section{Numerical simulations}
\label{section:test}
The historical data comes from the ICE (InterContinentalExchange) Market. The quotations begin in september 8th, 2003 and end in april 5th, 2007. 9 contracts for natural gas prices and 15 for crude oil are quoted for different maturity months.

\begin{figure}[htbp]
\begin{center}
\includegraphics[height=6cm,width=6cm]{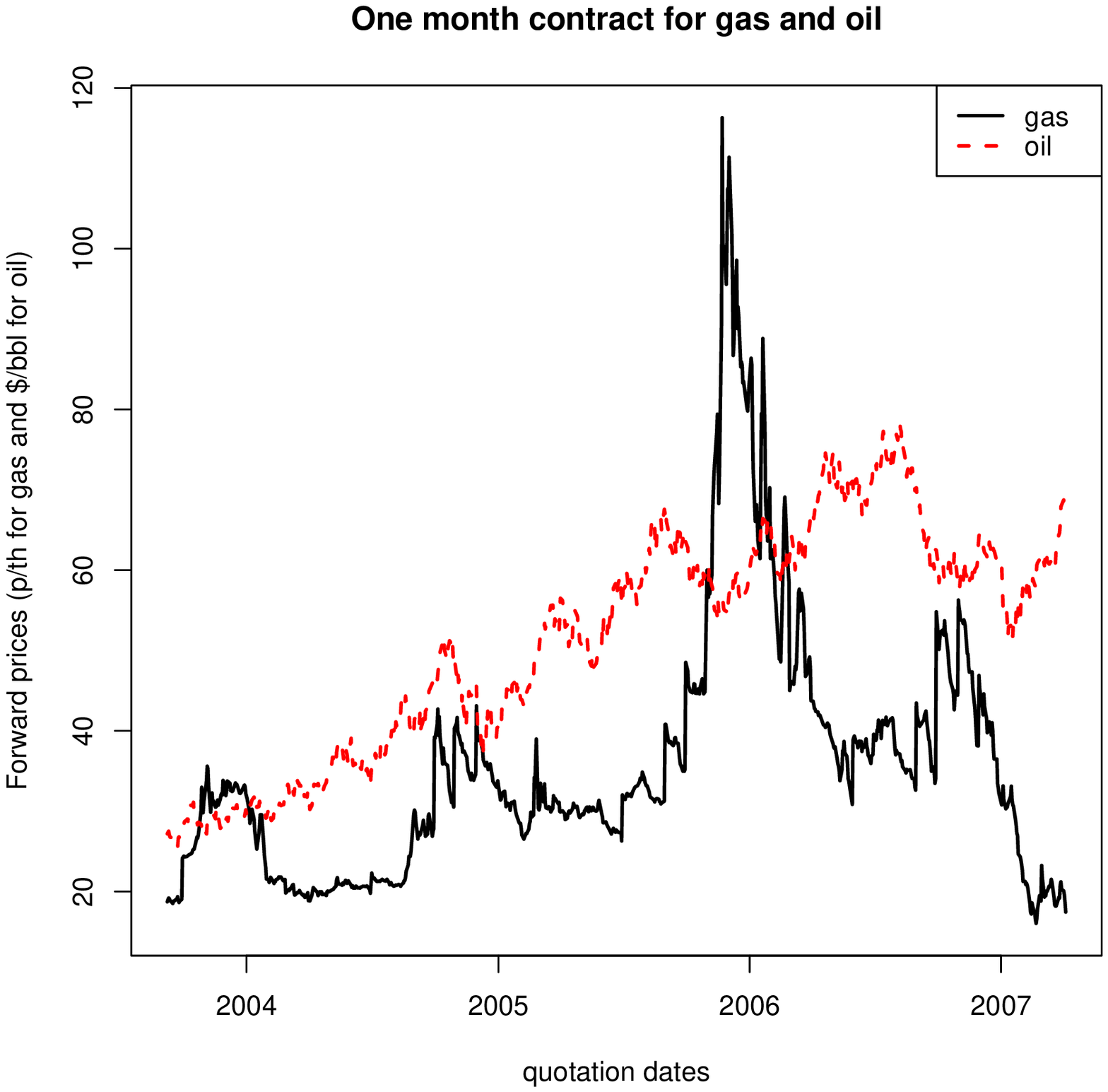}
\includegraphics[height=6cm,width=6cm]{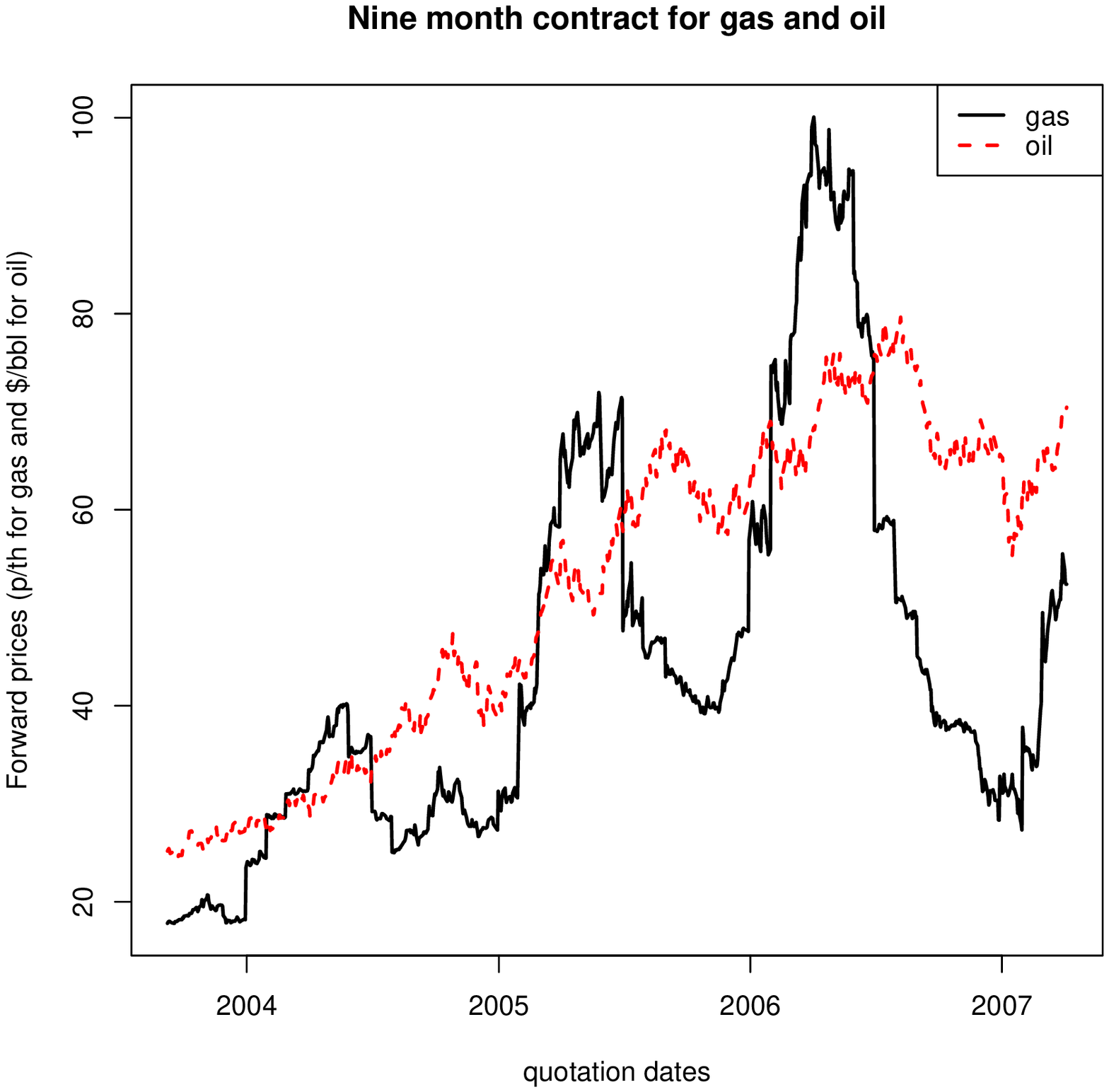}
\end{center}
\caption{Forward contracts delivering one unit of natural gas or crude oil in one month or nine months (p/th for natural gas and \$/bbl for crude oil).\newline
\small{ \textit{Despite the seasonality of gas, natural gas and crude oil prices seem to have a common long term tendency. }	}	\label{GasCrudeTenor}}	\end{figure}

The Figure \ref{GasCrudeTenor} represents the prices of the contract delivering one unit of gas or one unit of oil the next month (on the right hand side) and nine months later (on the left hand side). Even if the gas time series is seasonal, we notice that natural gas and crude oil prices have the same long term tendency.

\begin{figure}[htbp]
\begin{center}
\includegraphics[height=6cm,width=6cm]{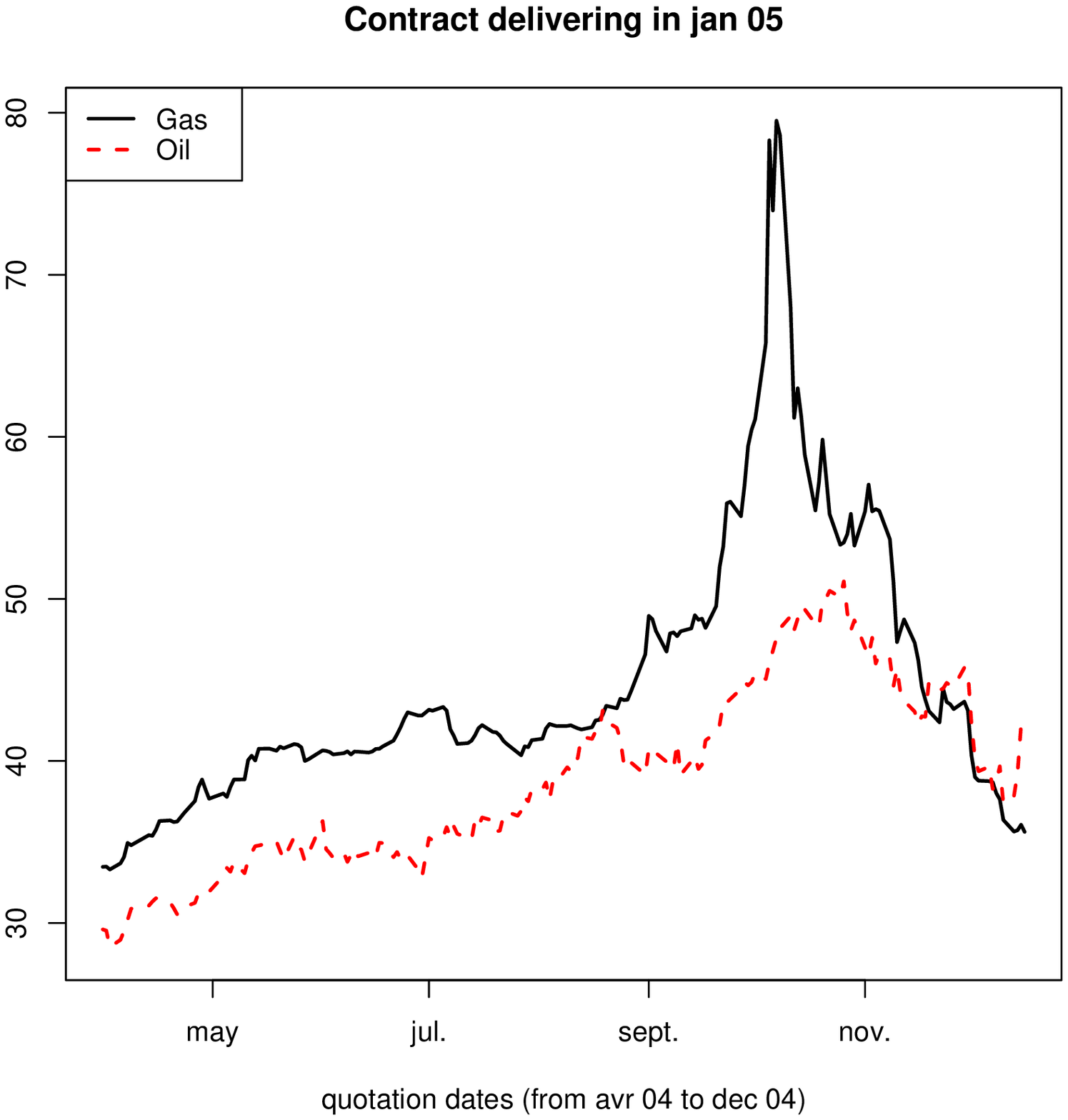}
\includegraphics[height=6cm,width=6cm]{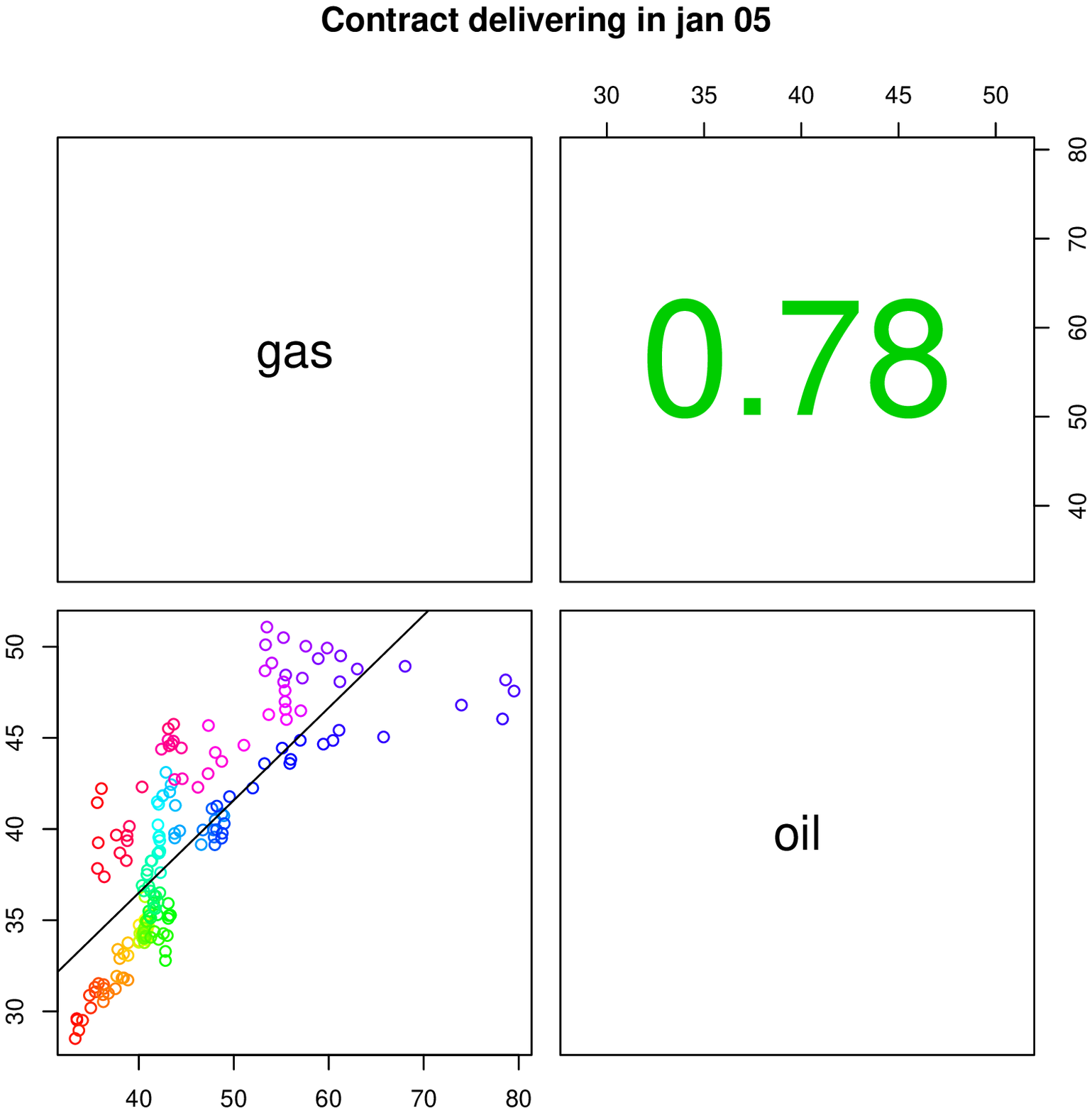}
\includegraphics[height=6cm,width=6cm]{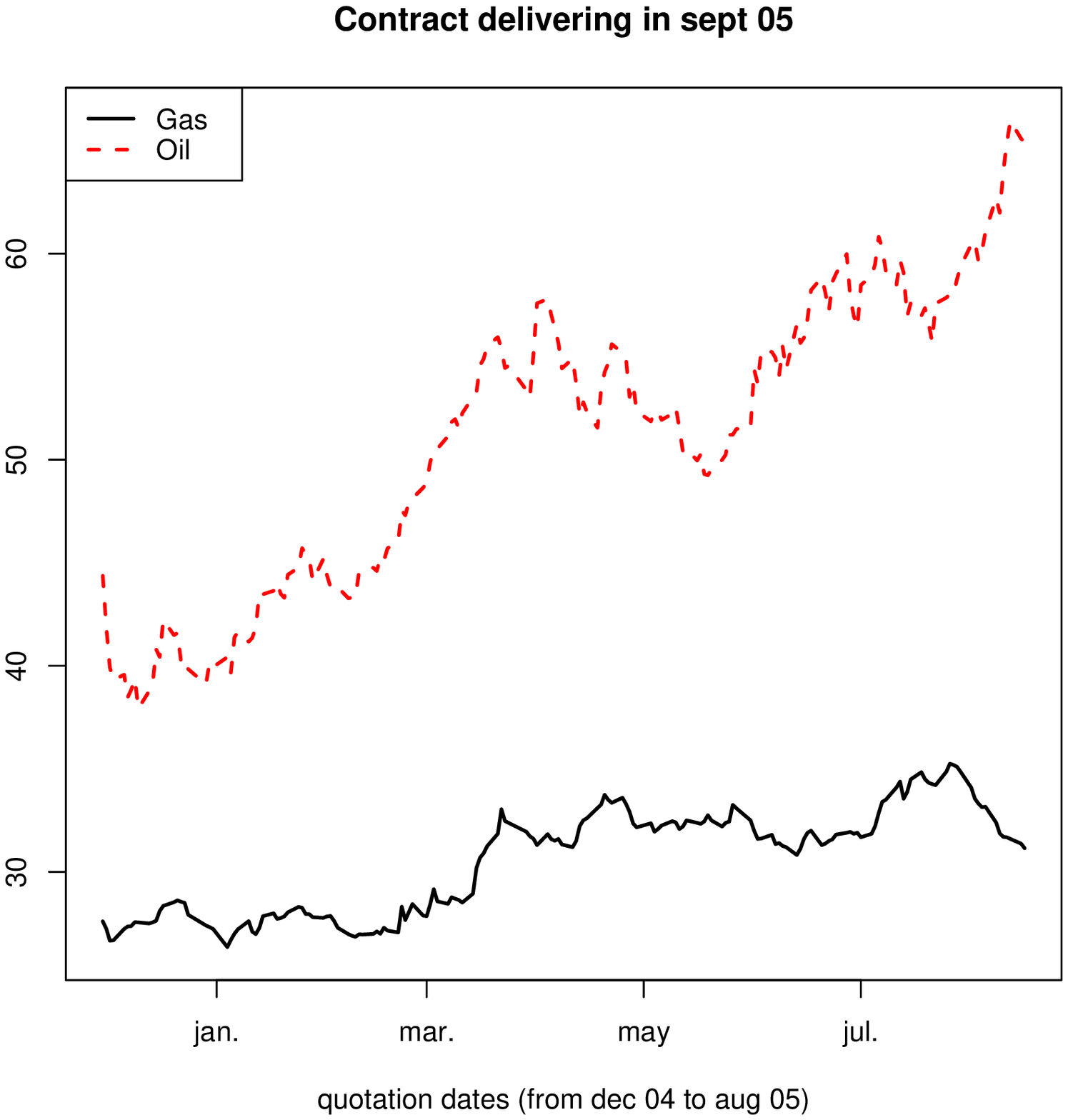}
\includegraphics[height=6cm,width=6cm]{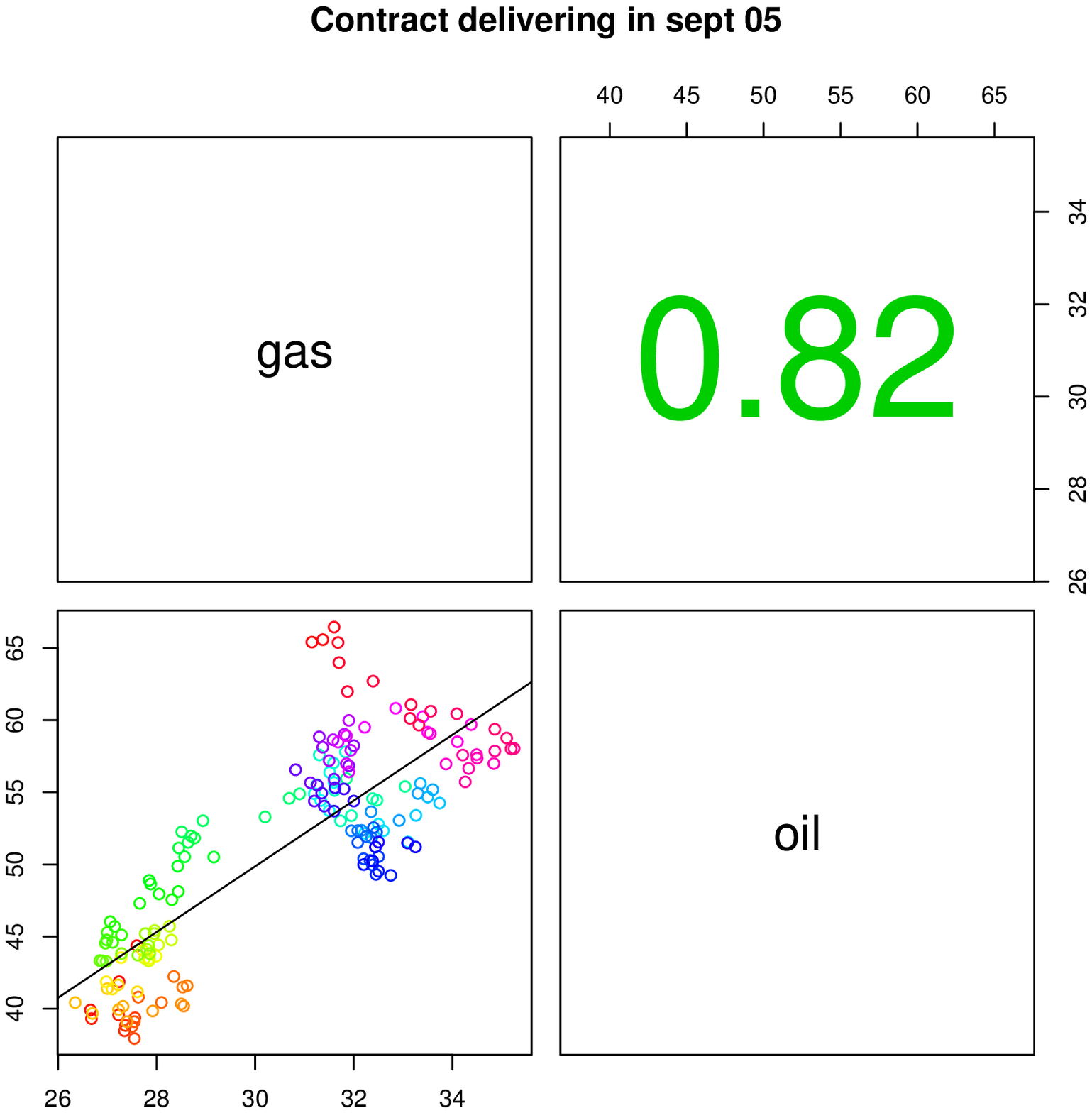}
\end{center}
\caption{Contracts delivering in January and September 2005 (p/th for natural gas and \$/bbl for crude oil).\newline
\small{ \textit{Natural gas and crude oil prices have a common long term tendency. }	}		\label{GasCrudeMat}}
\end{figure}

To avoid the seasonality effect, we can plot the gas and oil prices for a fixed delivering date (fixed maturity), see Figure \ref{GasCrudeMat}. This graph illustrates the long term tendency of natural gas and crude oil.

The first three steps of calibration allow us to obtain the following parameters (for $£N^g=N^c=3$):
\begin{center}
\begin{tabular}{|c|c|c|c|}
	\hline
	$\tau_1^g$ & $\tau_2^g$ & $\tau_1^c$ &$\tau_2^c$ \\
\hline 0.736 year & 0.086 year & 3.761 years & 0.138 year \\
	\hline
	\end{tabular}
\end{center}
The parameters $(\tau_1^e)_e$ convey the characteristic time of a short term shock. The impact of a shock on crude oil prices will last during a longer period than a shock on natural gas prices (i.e. 3.761 years $>$ 0.736 year). 

\begin{figure}[htbp]
\begin{center}
\includegraphics[height=6cm,width=6cm]{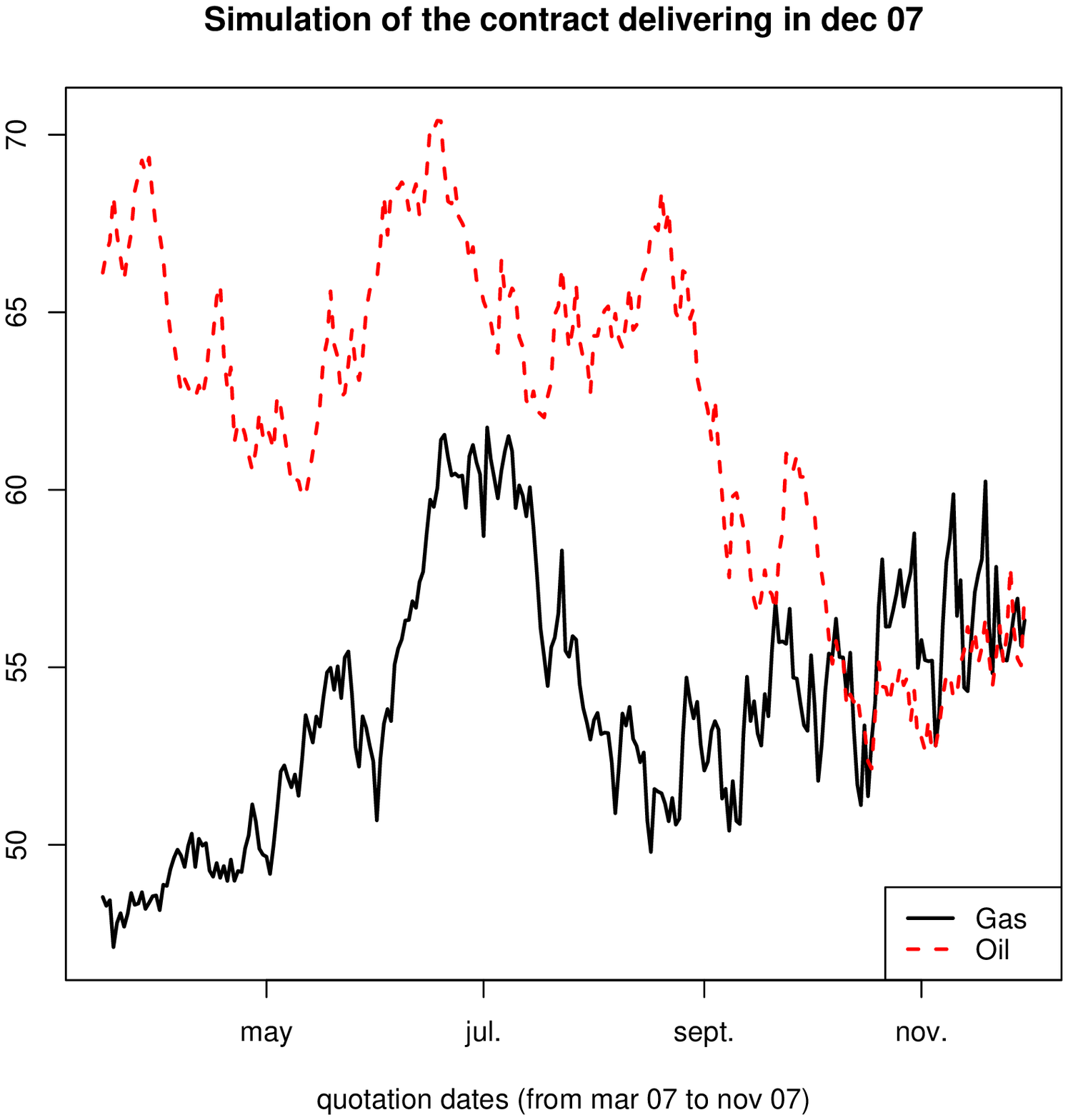}
\includegraphics[height=6cm,width=6cm]{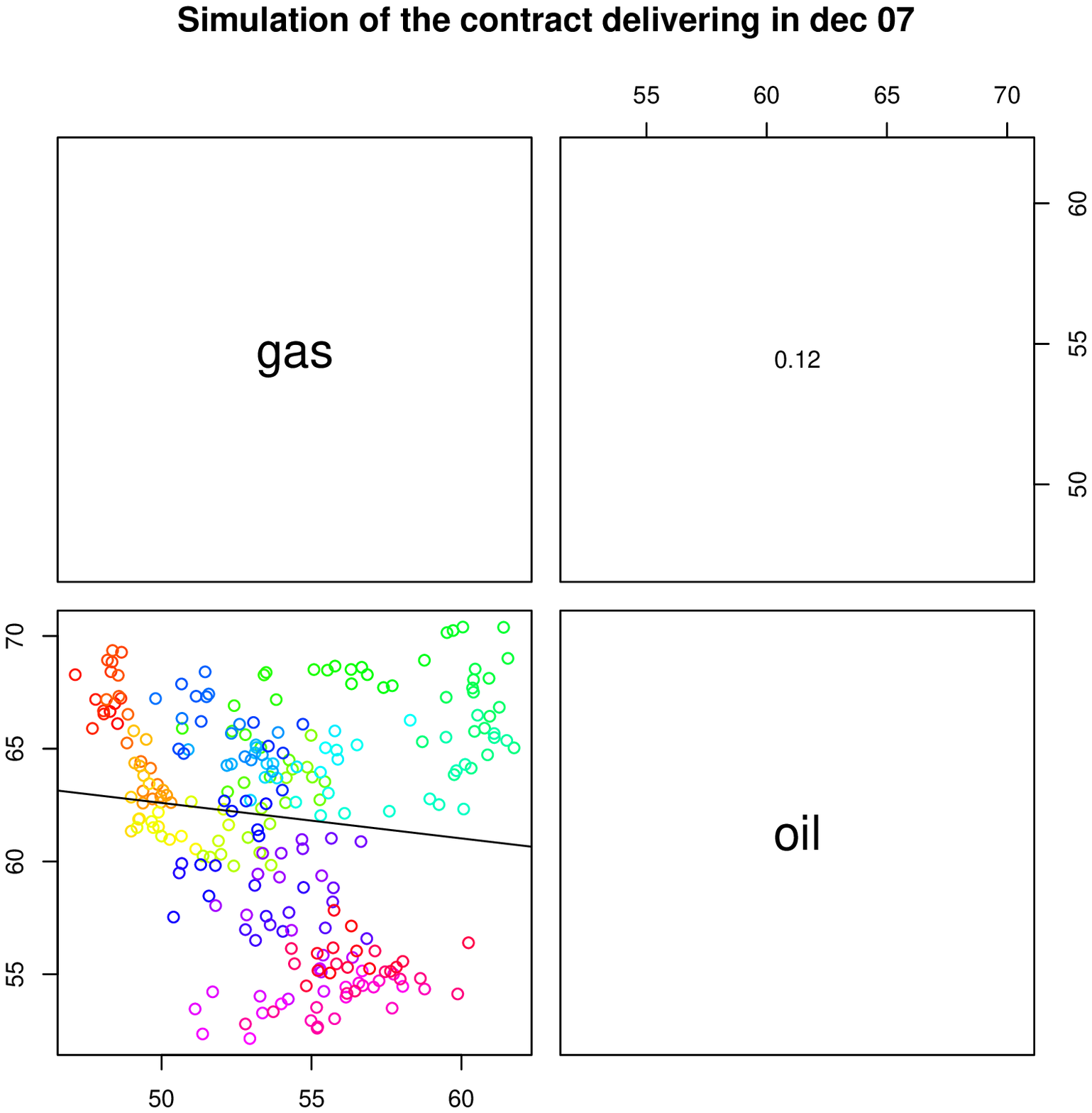}
\end{center}
\caption{Simulation for the contract delivering in December 2007 using Brooks' model.\newline
\small{ \textit{Natural gas and crude oil prices have no common long term tendency. }	}		\label{BrooksSimu}}
\end{figure}

Using Brooks' model with a simple correlation between Brownian motions to link energies, we can obtain such prices' simulation as in Figure \ref{BrooksSimu}.

\begin{figure}[htbp]
\begin{center}
\includegraphics[height=6cm,width=6cm]{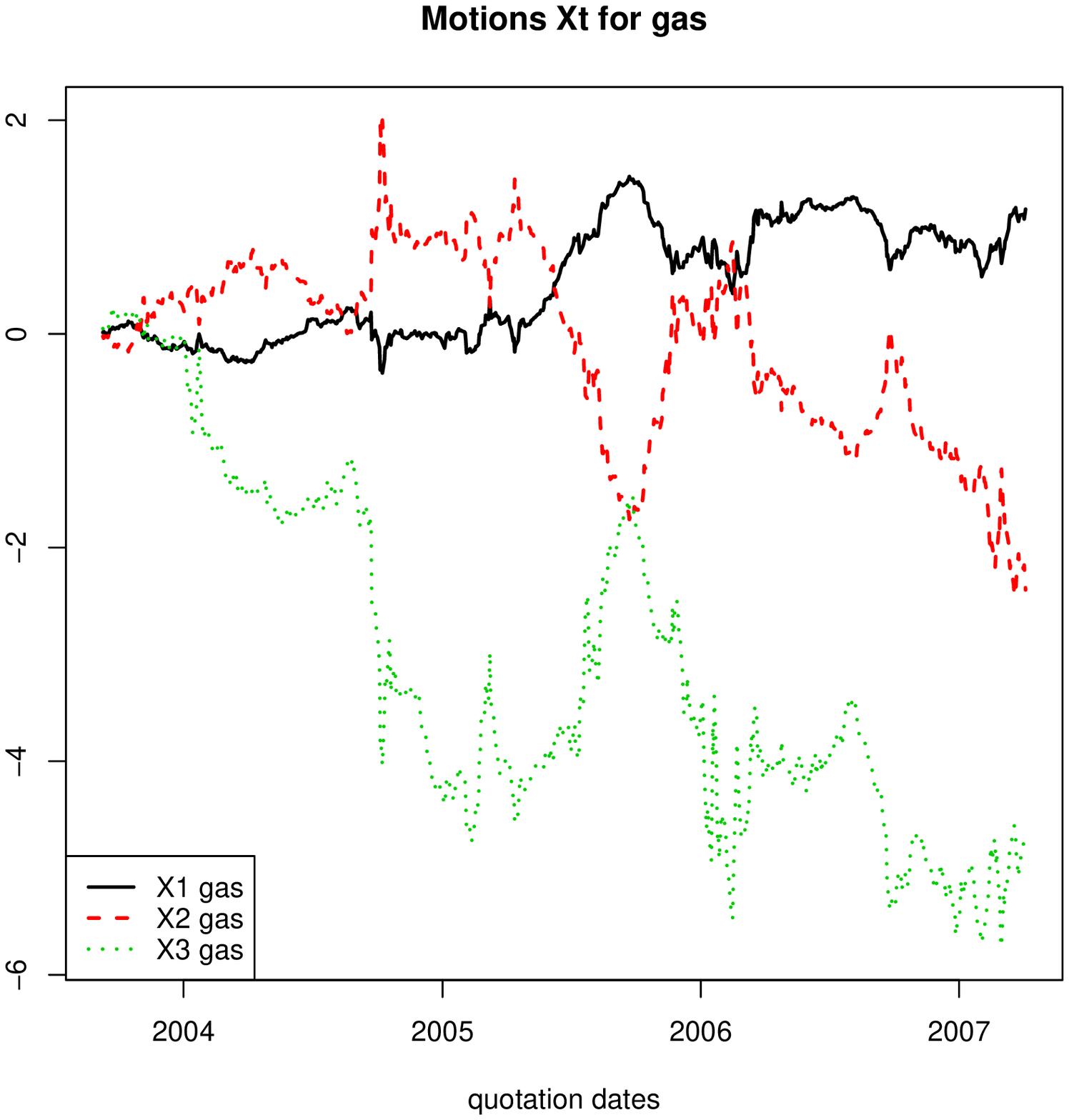}
\includegraphics[height=6cm,width=6cm]{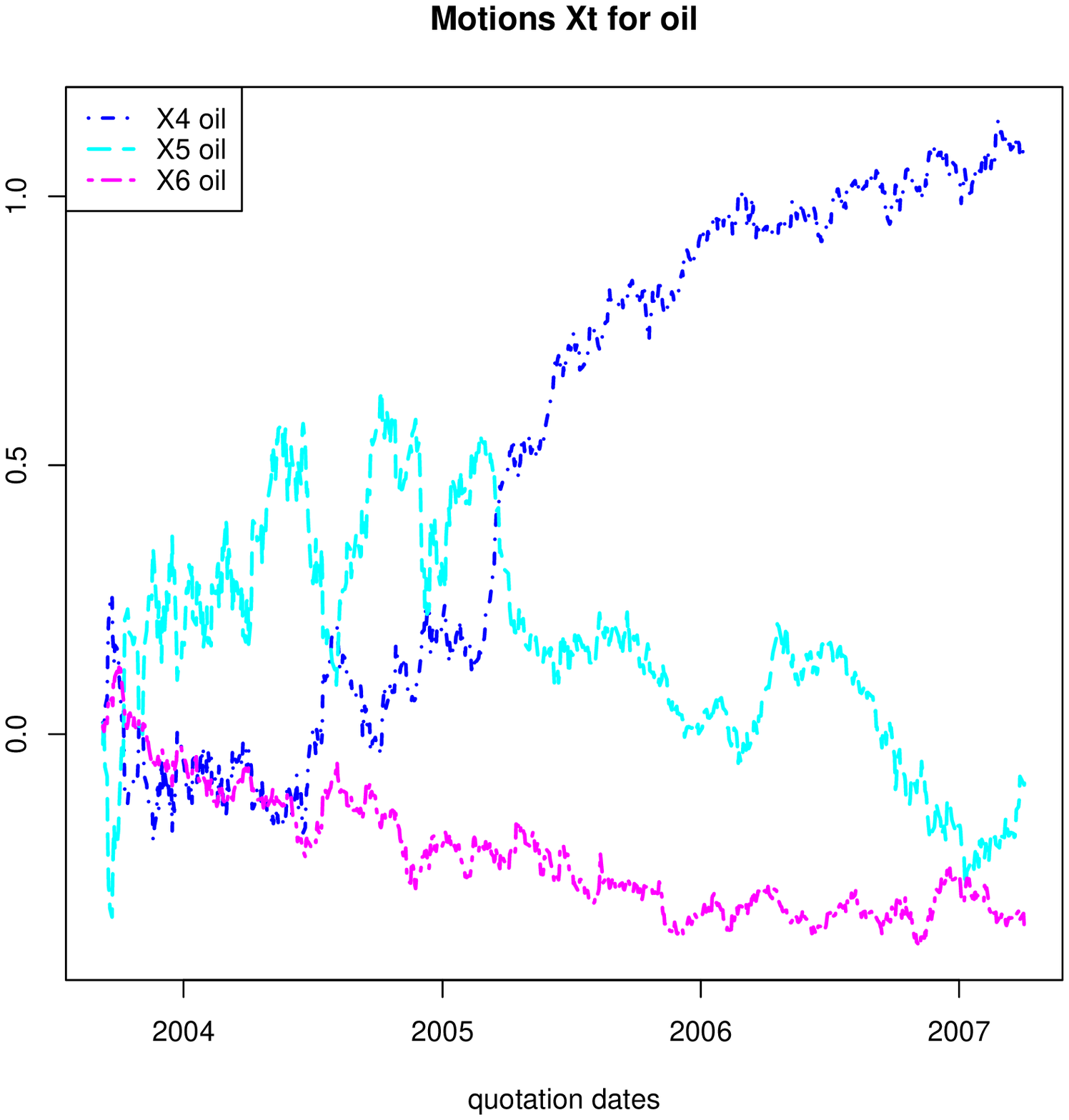}
\end{center}
\caption{Motions $(X_t)_t$ for natural gas and crude oil.\newline
\small{ \textit{According to the Phillipss-Ouliaris cointegration tests, the motions $(X_t)_t$ are cointegrated. }	}		
\label{Xt}}
\end{figure}

From the market data and the previous values of $\tau_i^e$, we obtain the motions $X_t$ (see Figure \ref{Xt}).

\begin{figure}[htbp]
\begin{center}
\includegraphics[height=6cm,width=6cm]{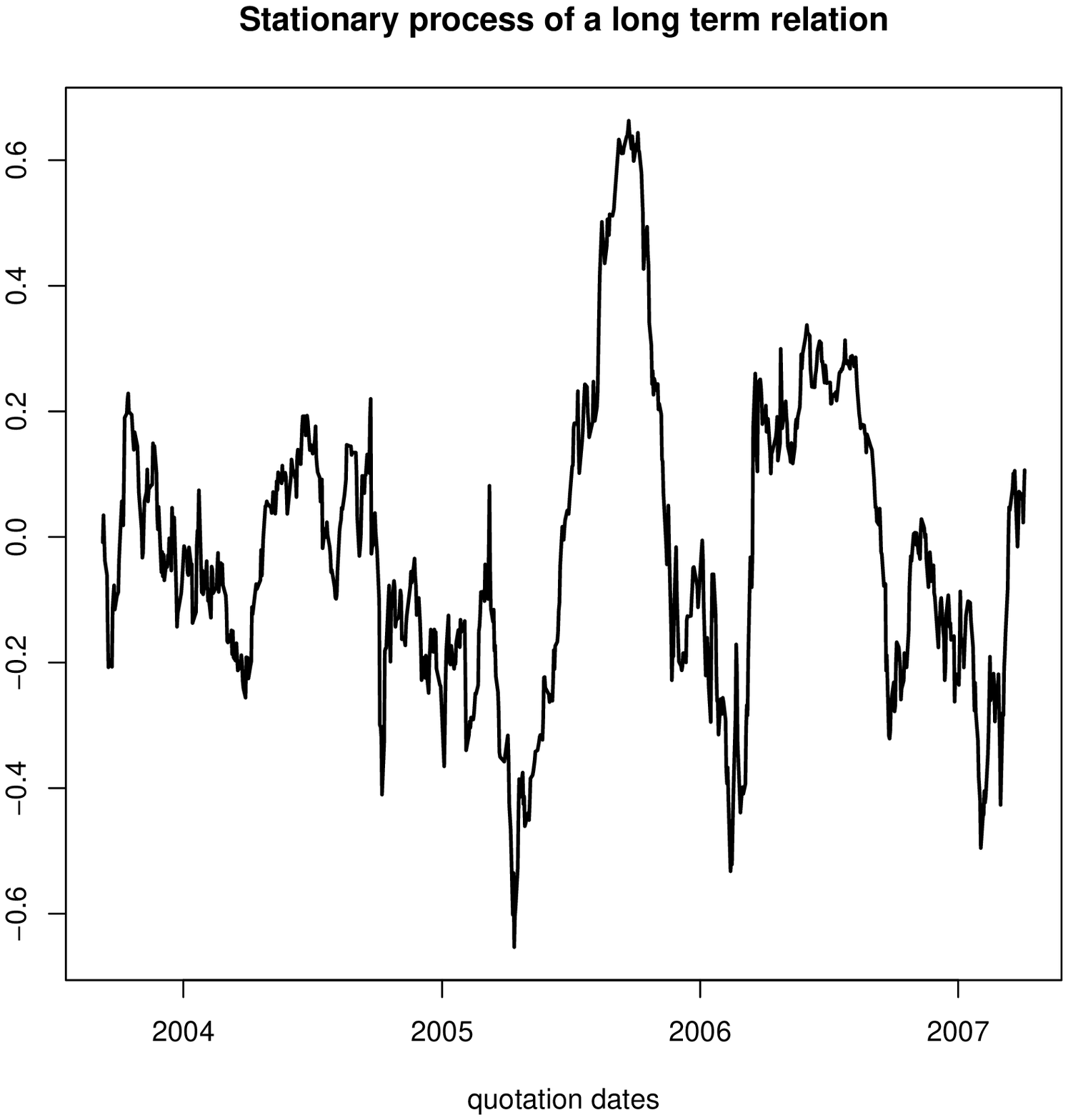}
\end{center}
\caption{Stationary process from one of the long term relations binding the motions $X_t$.\newline
\small{ \textit{The long term relation presented here is $X_t^1 - 0.97 X_t^4 = u_t$ where $(u_t)$ is a stationnay process. }	}	
\label{StationaryLT}}
\end{figure}

Figure \ref{StationaryLT} shows an example of stationary process resulting from one of the long term relations binding the motions $(X_t)_t$.

\begin{figure}[htbp]
\begin{center}
\includegraphics[height=6cm,width=6cm]{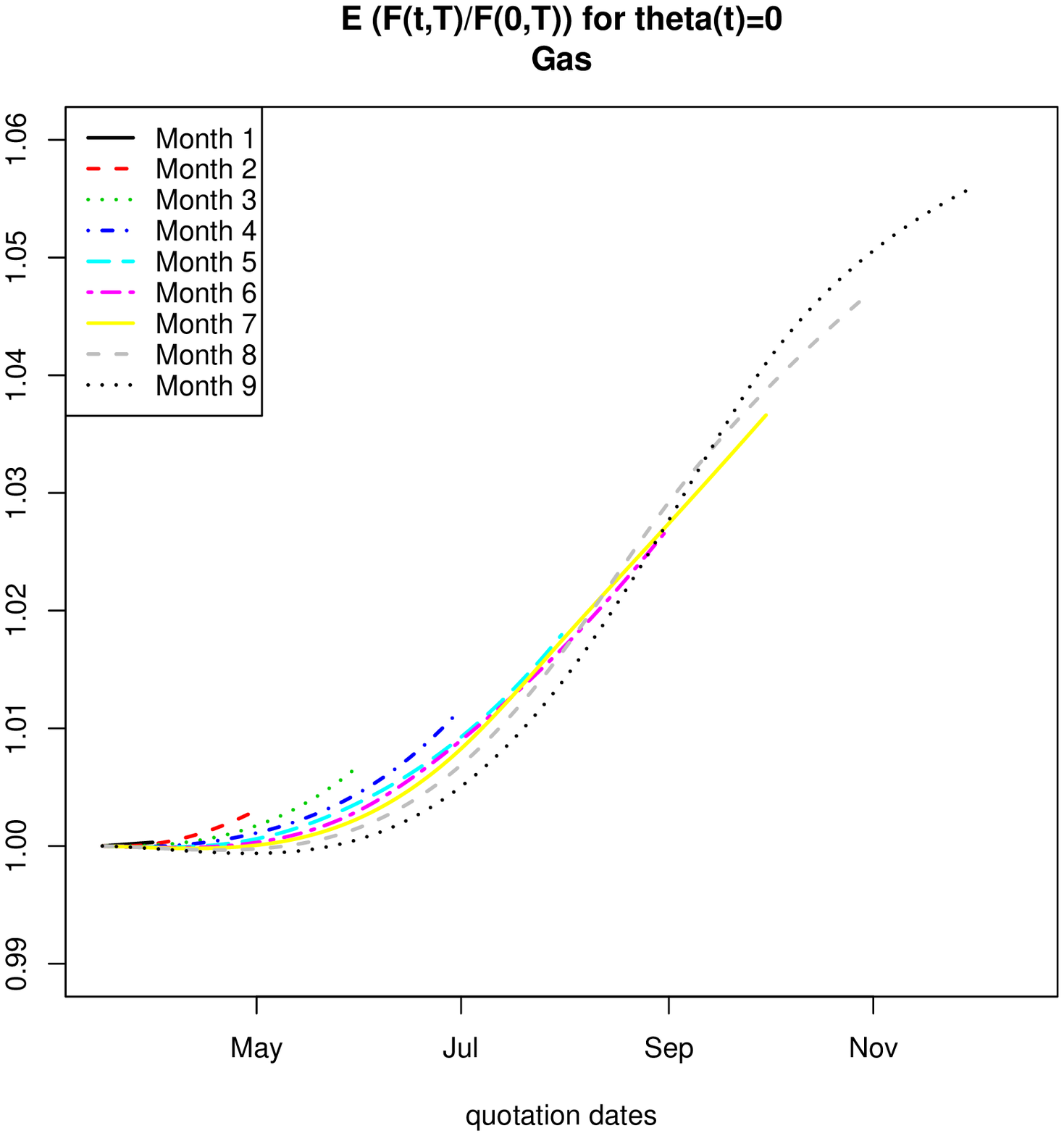}
\includegraphics[height=6cm,width=6cm]{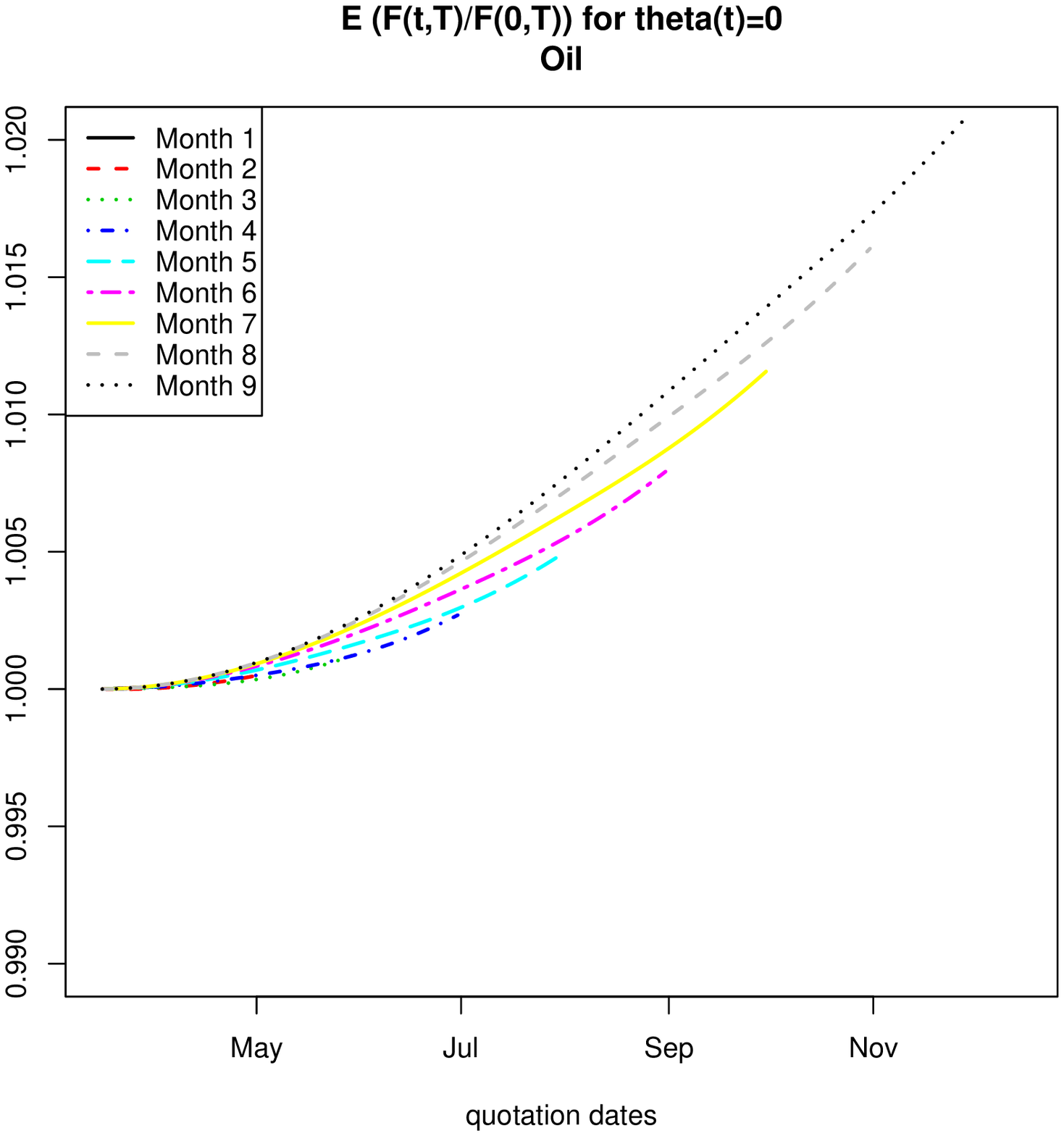}
\includegraphics[height=6cm,width=6cm]{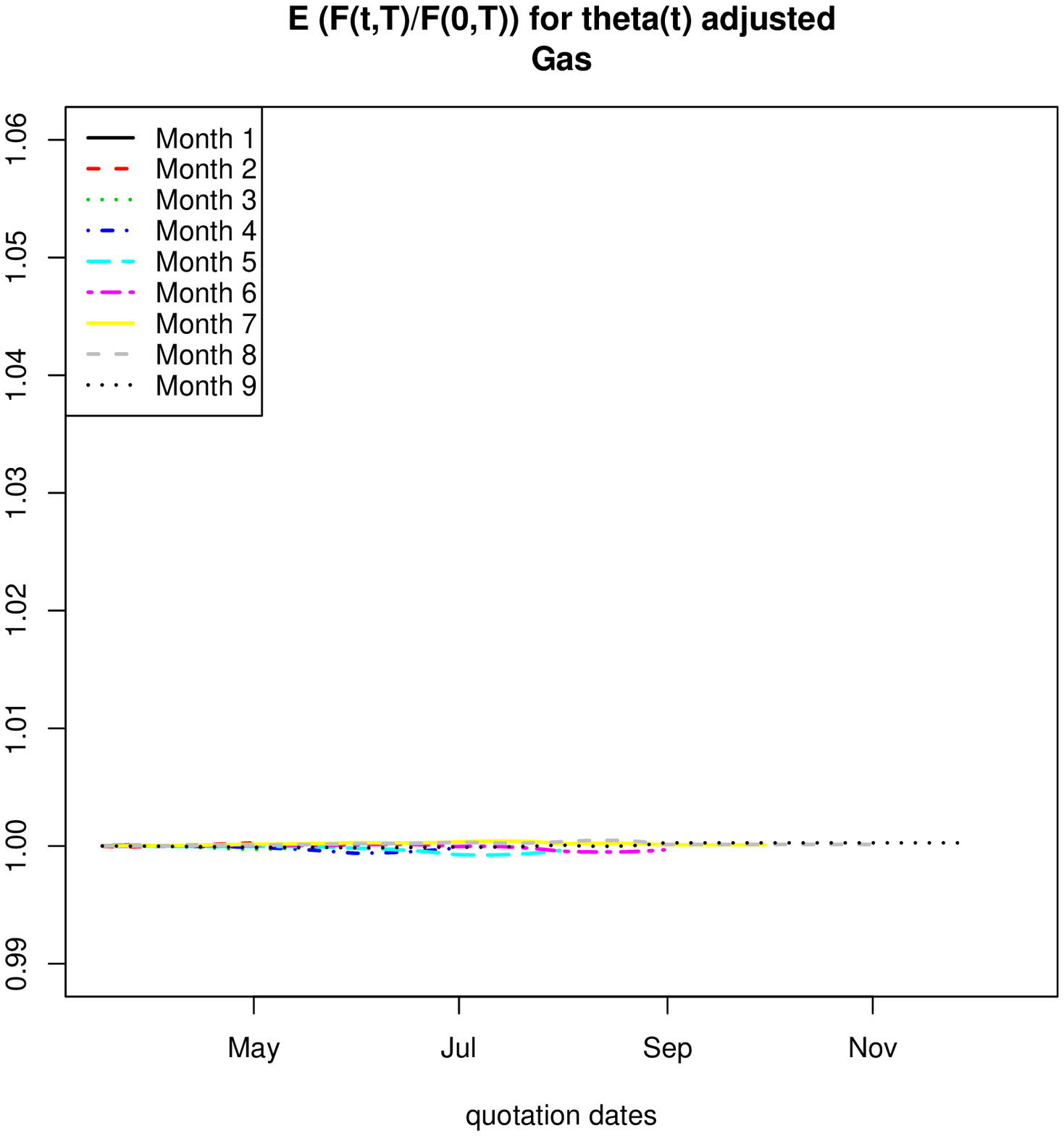}
\includegraphics[height=6cm,width=6cm]{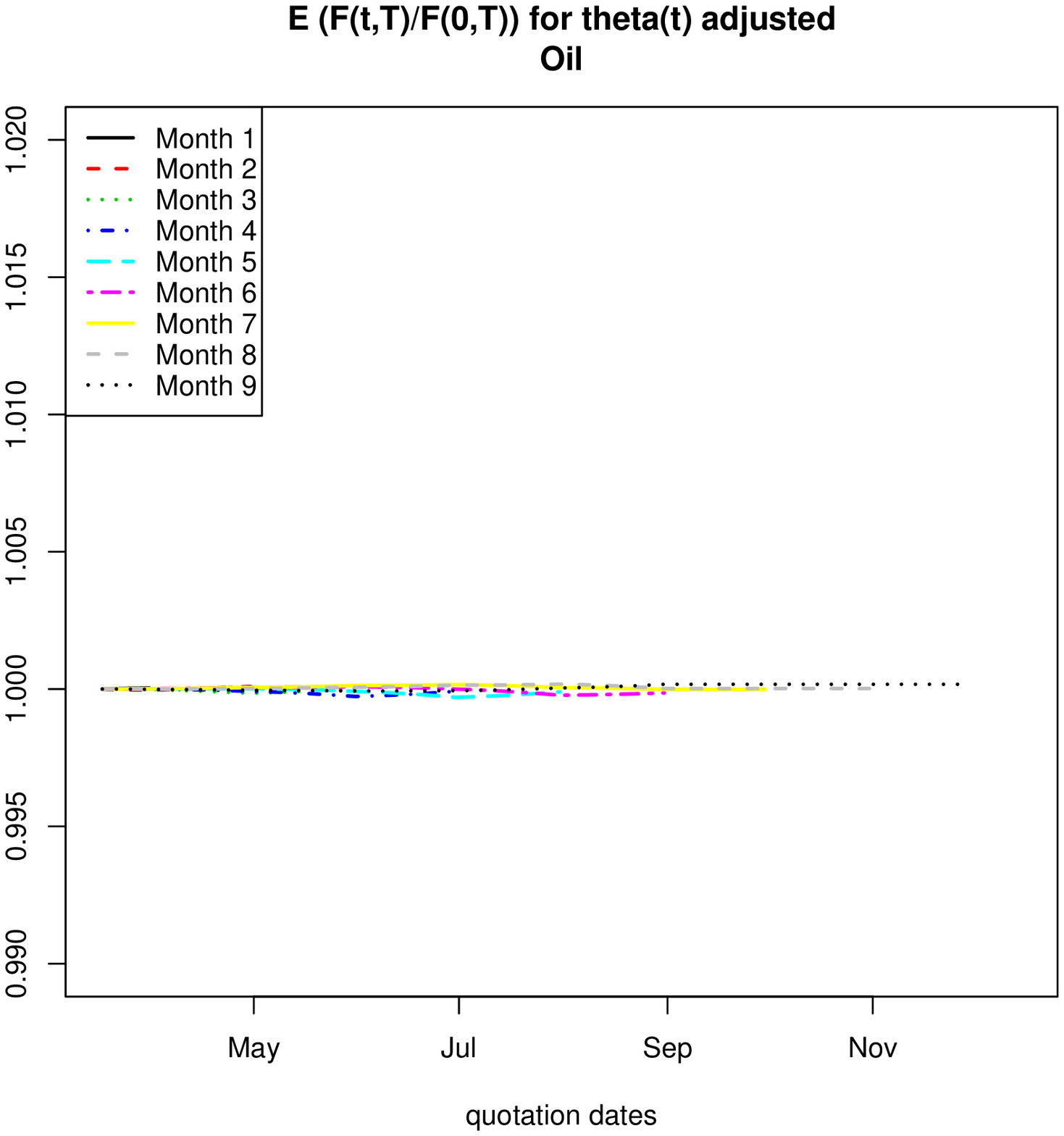}
\end{center}
\caption{Thanks to the parameter $\theta_t$, prices' simulations are centred on the initial forward curve.\newline
\small{ \textit{The ratio of expectations must be 1. In the first case ($\forall t, \ \theta_t=0$), we notice a maximum difference with 1 of 5.6\% for gas and 2.1\% for oil. In the second case ($\theta_t$ is now adjusted), the maximum deviation from 1 is 0.08\% for gas and 0.03\% for oil. }	}	\label{Theta}}
\end{figure}

\begin{figure}[htbp]
\begin{center}
\includegraphics[height=6cm,width=6cm]{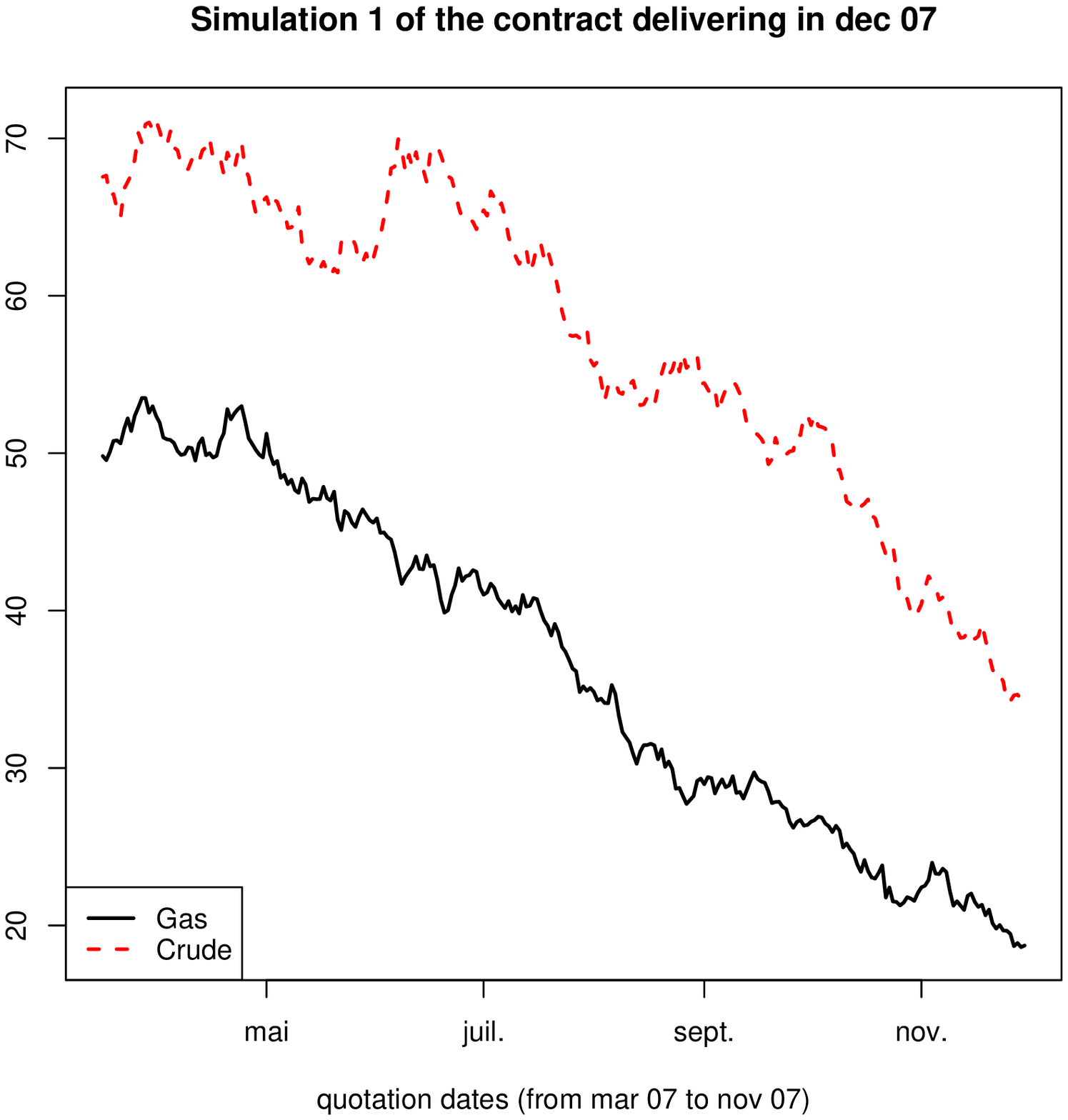}
\includegraphics[height=6cm,width=6cm]{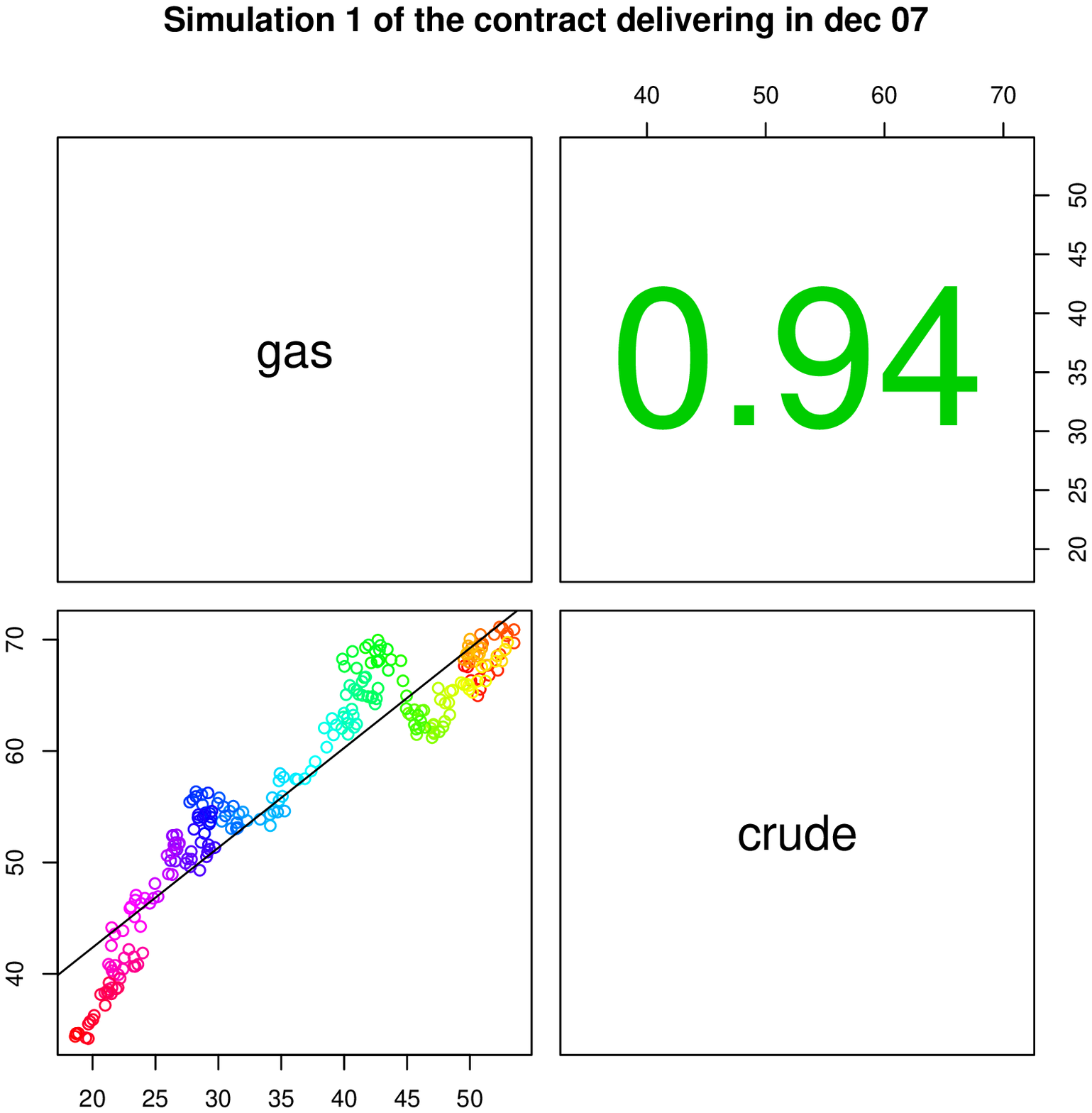}
\includegraphics[height=6cm,width=6cm]{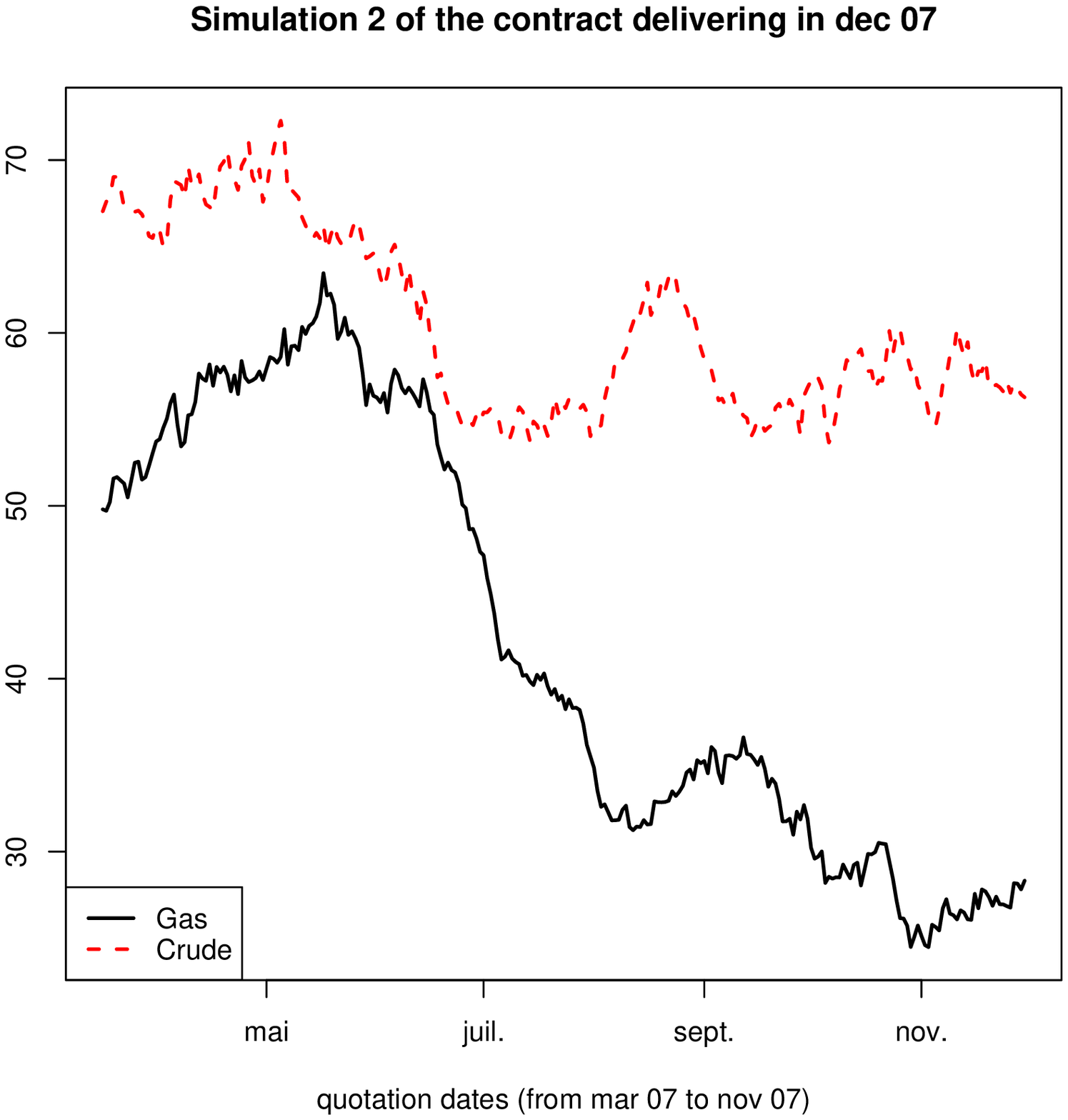}
\includegraphics[height=6cm,width=6cm]{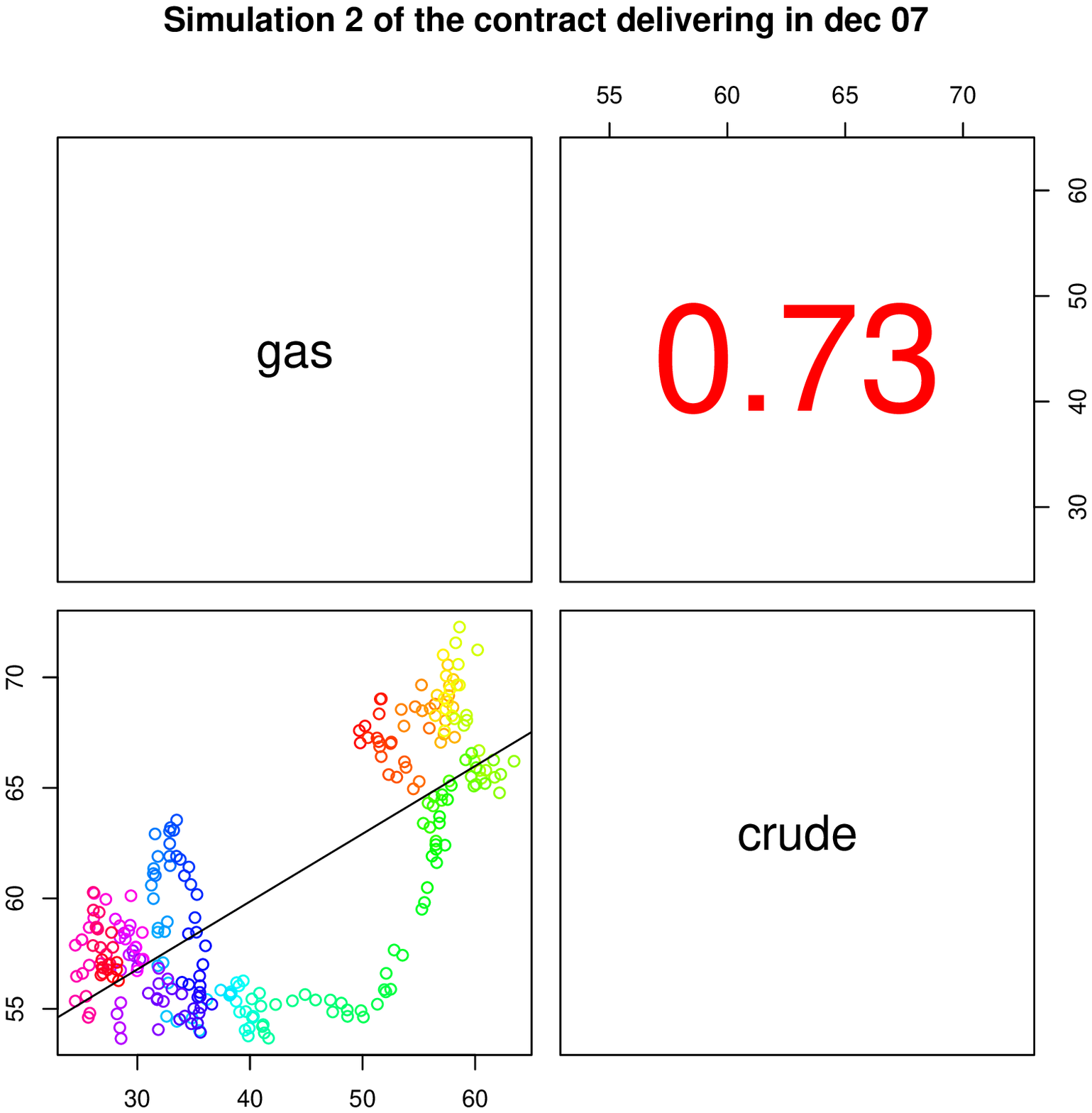}
\includegraphics[height=6cm,width=6cm]{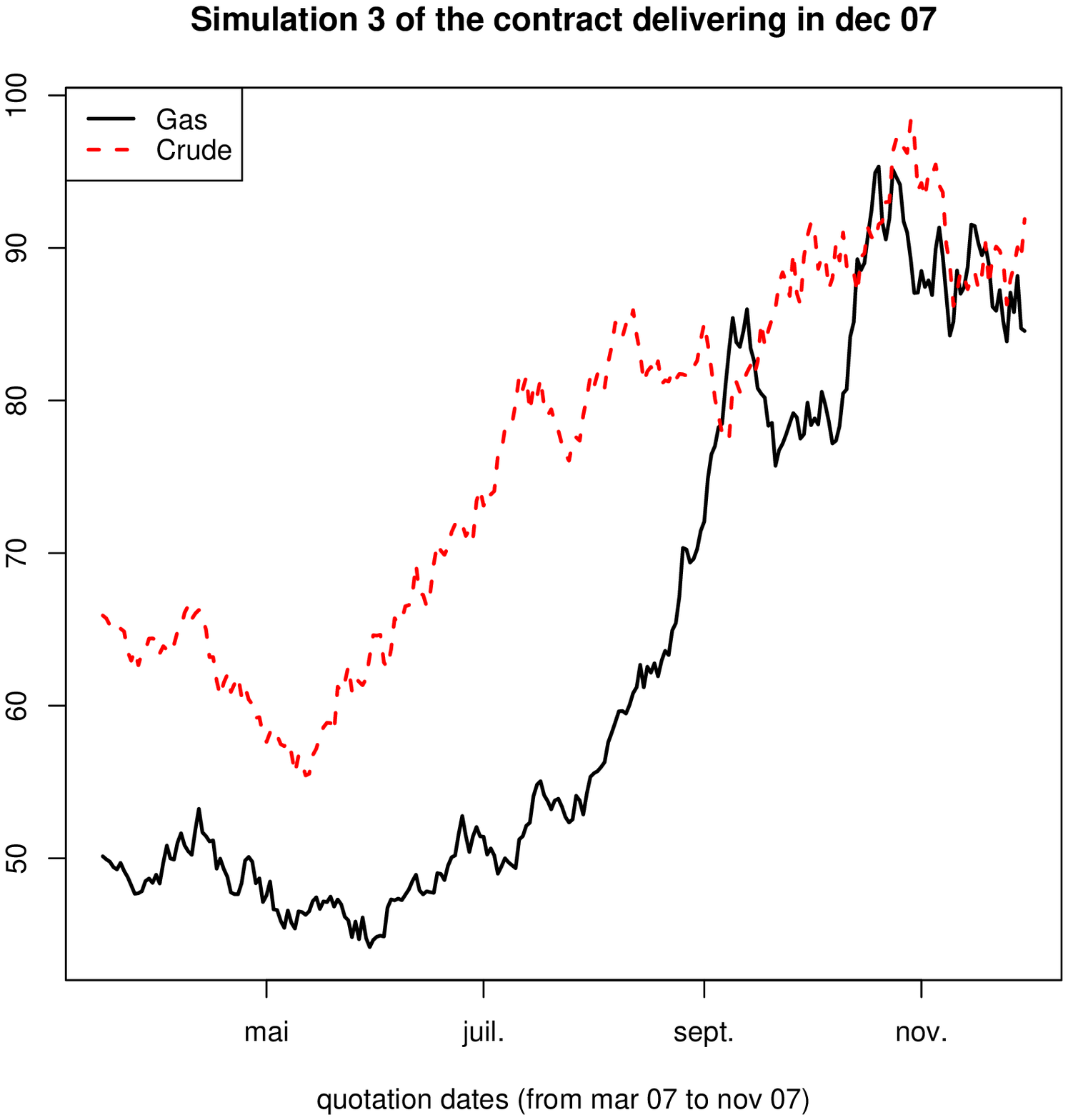}
\includegraphics[height=6cm,width=6cm]{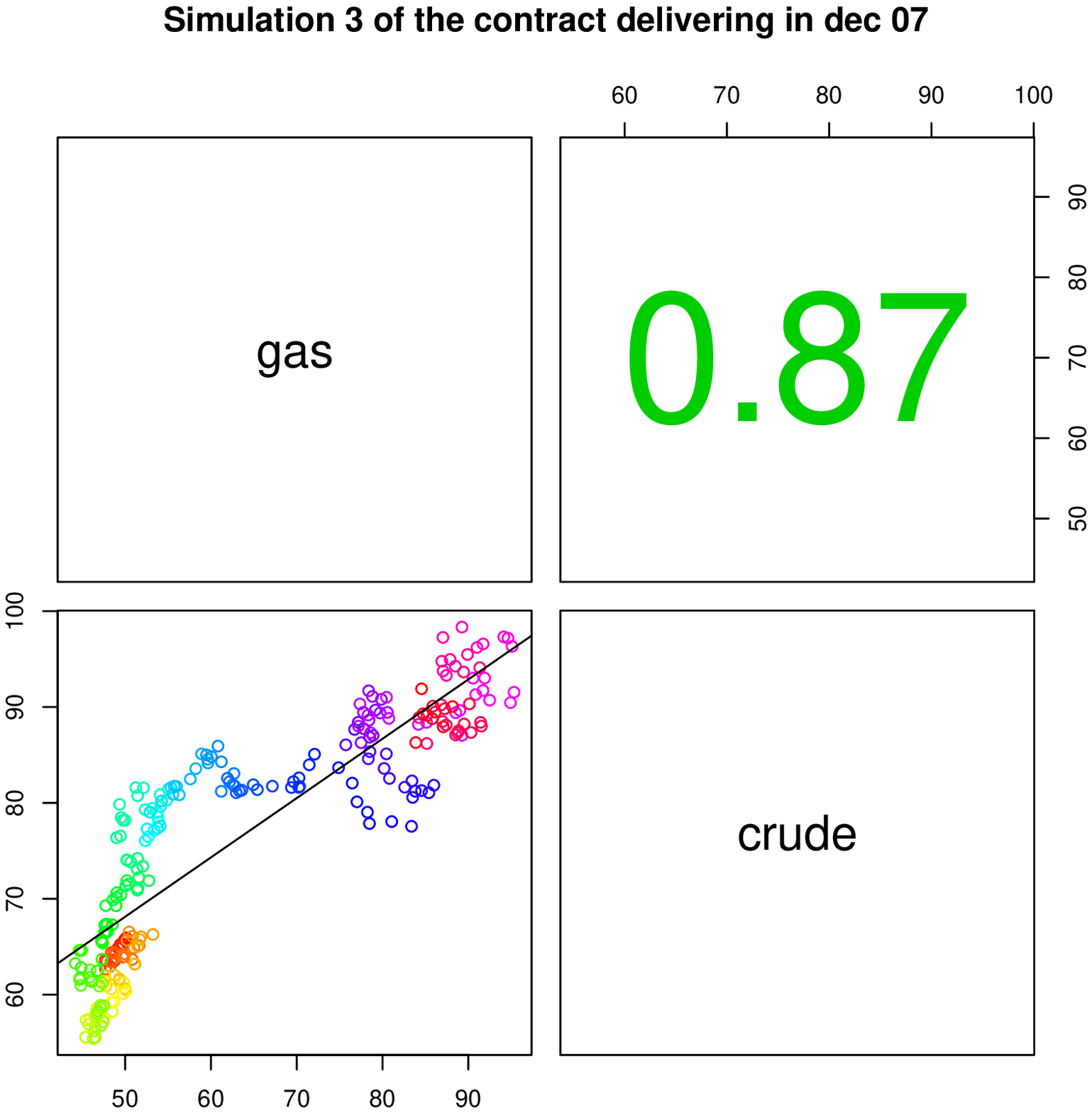}
\end{center}
\caption{Three examples of simulations for contracts delivering in one month.
\newline
\small{ \textit{The model conveys the long term relation between natural gas and crude oil contracts. }}
\label{GasCrudeS}}
\end{figure}

Cointegration tests (Phillips-Ouliaris) allow us to reject the null hypothesis of non-cointegration. This justifies the use a Vectorial Error Correction Model to model the motions $X_t$. With significant linear regressions (according to the BIC), we find the following matrices for $\Pi$ and $\Sigma \Sigma^*$:
\begin{align*}
\Pi &= \left(
\begin{array}{cccccc}
-0.017  & 0 &  0 &  0.019 & 0 &  0 \\
 0  & -0.005 &  0.009 & -0.027  & 0 & -0.162 \\
 0  &  0 & -0.012 & 0 & 0 &  0.174 \\
 0  &  0 & 0 & -0.009 & 0 & -0.030 \\
 0  &  0 & 0 & 0.015 & 0.008 &  0.046 \\
 0  &  0 & 0 & -0.017 & 0.019 & -0.052
 \end{array}
\right),\\
\Sigma \Sigma^* &= \left(
\begin{array}{cccccc}
  0.00158 & -0.00323 &  0.00386 & -0.00001 &  0.00006 &  0.00003 \\
 -0.00323 &  0.00812 & -0.00958 & -0.00007 &  0.00007 & -0.00007 \\
  0.00386 & -0.00958 &  0.01740 &  0.00006 & -0.00003 &  0.00011 \\
 -0.00001 & -0.00007 &  0.00006 &  0.00045 & -0.00052 &  0.00010 \\
  0.00006 &  0.00007 & -0.00003 & -0.00052 &  0.00096 & -0.00011 \\
  0.00003 & -0.00007 &  0.00011 &  0.00010 & -0.00011 &  0.00015	
  \end{array}
\right).
\end{align*}
All these parameters allow us to compute $\Eb_{\theta=0}^{\Pb} \left( \frac{F^e(t,T)}{F^e(0,T)} \right)$ in order to centre the simulations on the initial forward curve. Figure \ref{Theta} shows this result for fixed $(T-t)$.
We now show some examples of price simulations using this model. To simulate one scenario, we use an Euler scheme. The prices are diffused from the initial forward curves of March 15th, 2007. We draw the simulations for a fixed maturity so as to favour the long term relation expressed by the model, over the seasonality of gas. Three simulations are shown on Figure \ref{GasCrudeS}. These simulations fit well the possible scenarios given by market data (see Figure \ref{GasCrudeMat}).


\section*{Conclusion}
In this work, we propose a continuous time model for natural gas and crude oil future markets, which brings together two key features of these energy markets. On the one hand, it conveys a long term dependence between gas and oil, as it has already been supported by several previous econometric studies. This dependence is achieved through a cointegration of motions driving the prices. On the other hand, the model is free of arbitrage and coincides with usual forward models under the risk neutral probabilities.  As a consequence, this model can be used for Value at Risk computations (or other risk management measurements) as well as for option pricing.\\
Based on market data related to future contracts, the calibration gives very good results. It would be presumably improved by incorporating derivatives prices in the calibration set.\\ 
We have presented a version of the model without seasonality effect (especially regarding gas). We mention finally that it is easy to incorporate seasonality in the gas volatility.

\bibliographystyle{elsart-harv.bst}
\bibliography{biblio}


\section*{Appendix}
We first recall the exponential form of the forward price: 
\begin{align*}
  F^e(t,T)=F^e(0,T)\exp&\left(\int_0^t \sigma^e (T-s) \theta'_s \dd s\right)\\
&\exp\left(\int_0^t \sigma^e (T-s) \dd \tilde X_s -\frac 12 \langle\sigma^e (T-.) \tilde X_. \rangle _t\right).
\end{align*}
The above quadratic variation equals $\int_0^t |\sigma^e(T-s)\Sigma|^2 \dd s$. Since $\tilde X$ does not depend on $\theta$, the assertion \eqref{eq:factorisation} is proven.

Let us prove the following formula
 \begin{align}
\label{eq:closed:formula}
\Eb_{\theta=0}^{\Pb} \left( \frac{F^e(t,T)}{F^e(0,T)} \right)  =  
&\exp \bigg[ \int_0^t \left( \frac{1}{2} \left| \int_s^t \sigma^e(T-u)  \Pi e^{(u-s)\Pi} \dd u \Sigma \right|^2   \right. \\
   & +  \left. \sigma^e(T-s) \Sigma\Sigma^* \int_s^t e^{(u-s)\Pi^*} \Pi^*  (\sigma^e(T-u))^* \dd u \right)  \dd s \bigg]\nonumber.
\end{align}
Set $I=\int_0^t \sigma^e (T-s) \dd \tilde X_s$; then, one has 
\begin{equation}
  \label{eq:I}
  \Eb_{\theta=0}^{\Pb} \left( \frac{F^e(t,T)}{F^e(0,T)} \right)=\Eb^{\Pb}\left(
\exp(I)\right) \exp\left(-\frac 12 \int_0^t |\sigma^e(T-s)\Sigma|^2 \dd s\right).
\end{equation}
Our methodology to compute $\Eb^{\Pb}\left(\exp(I)\right)$ consists in showing that $I$ is a centred Gaussian variable with a variance $v$ to calculate; it follows that $\Eb^{\Pb}\left(\exp(I)\right)=\exp(v^2/2)$. The process $\tilde X$ given in Equation \eqref{eq:model:bis} is a generalized Ornstein-Uhlenbeck process. Its solution is explicit and is given by:
\begin{equation}
  \label{eq:x}
  \tilde X_t=\int_0^t e^{(t-s)\Pi}\Sigma \dd W_s.
\end{equation}
Then, using again Equation \eqref{eq:model:bis}, we obtain
\begin{align*}
  I&=\int_0^t \sigma^e (T-s) \Pi \tilde X_s \dd s + \int_0^t \sigma^e (T-s) \Sigma \dd W_s\\
&=\int_0^t \sigma^e (T-s) \Pi \int_0^s e^{(s-u)\Pi}\Sigma \dd W_u \dd s+\int_0^t \sigma^e (T-s) \Sigma \dd W_s\\
&=\int_0^t \left[\int_s^t \sigma^e (T-u) \Pi e^{(u-s)\Pi}\Sigma \dd u +\sigma^e (T-s) \Sigma \right] \dd W_s.
\end{align*}
Thus, $I$ is a Wiener stochastic integral, whose is centred and whose variance equals $\int_0^t \left|\int_s^t \sigma^e (T-u) \Pi e^{(u-s)\Pi}\Sigma \dd u +\sigma^e (T-s) \Sigma \right|^2 \dd s$. Changing in \eqref{eq:I} and making simplifications lead to \eqref{eq:closed:formula}.

%

\end{document}